\documentclass[pdflatex, sn-mathphys-num]{sn-jnl}

\usepackage{xstring} 
\usepackage{xspace}
\usepackage{units}
\usepackage{verbatim}
\usepackage{amsmath}
\usepackage{amssymb}
\usepackage{ifthen} 
\usepackage{xifthen} 
\usepackage[T1]{fontenc}
\usepackage{array}   
\usepackage{lineno}
\usepackage{graphicx}
\usepackage{float}

\newcolumntype{L}{>{$}l<{$}} 
\newcolumntype{R}{>{$}r<{$}} 

\newcommand{\teff}{\ensuremath{T_{\mathrm{eff}}}\xspace}

\newcommand{\gcc}{g\,cm$^{-3}$\xspace}

\newcommand{\kepler}{\textit{Kepler}\xspace}
\newcommand{\ktwo}{\textit{K2}\xspace}
\newcommand{\spitzer}{\textit{Spitzer}\xspace}
\newcommand{\tess}{\textit{TESS}\xspace}

\newcommand\arcmin{\mbox{$^\prime$}}%
\newcommand\arcsec{\mbox{$^{\prime\prime}$}}%
\newcommand\micron{\mbox{$\mu$m}}%
\newcommand\ms{m~s$^{-1}$\xspace} 
\newcommand\nat{Nature}
\newcommand\aj{AJ}
\newcommand\apj{ApJ}

\newcommand\apjl{ApJL}
\newcommand\aap{A\&A}
\newcommand\mnras{MNRAS}
\newcommand\pasp{PASP}
\newcommand\pasj{PASJ}
\newcommand\psj{PSJ} 
\newcommand{\tc}[2][red]{\textcolor{#1}{\emph{#2}}}
\newcommand{\isr}[2]{\ensuremath{#1_\mathrm{#2}}\xspace}
\newcommand{\Zfree}{\ensuremath{Z_{\mathrm{free}}}\xspace} 
\newcommand{\Zfreestar}{\ensuremath{Z_{\mathrm{free}}^{*}}\xspace}
\newcommand{\zfree}[1]{\ensuremath{\ifthenelse{\isempty{#1}}{z_{\mathrm{free}}^{*}}{z_{\mathrm{free,#1}}}}\xspace}

\newcommand{\Me}{\ensuremath{M_{\oplus}}\xspace} 
\renewcommand{\Re}{\ensuremath{R_{\oplus}}\xspace}

\newcommand{\Msun}{\ensuremath{M_{\odot}}\xspace}

\newcommand{\sname}{V1298~Tau\xspace}
\newcommand{\age}{10--30~Myr}
\newcommand{\val}[1]{
\IfEqCase{#1}{
{mass-c}{4.7 \pm 0.6}
{mass-d}{6.0 \pm 0.7}
{mass-b}{13.1\pm 5.3}
{mass-e}{15.3\pm 4.2}
{radius-c}{5.08 \pm 0.37}
{radius-d}{6.53 \pm 0.42}
{radius-b}{9.41 \pm 0.57}
{radius-e}{10.17\pm 0.75}
{rho-c}{0.20 \pm 0.05}
{rho-d}{0.12 \pm 0.03}
{rho-b}{0.09 \pm 0.04}
{rho-e}{0.08 \pm 0.03}
{n-ground-based-transits}{\tc{43}} 
{delta-cd-per}{0.23}
{delta-db-per}{-2.67}
{delta-be-per}{0.82}
{ttv-super-per-cd-lin}{1834}
{ttv-super-per-db-lin}{451}
{ttv-super-per-be-lin}{2953}
{n-cd}{53}
{ttv-cd-bestfit-resid-min-std}{11}
{ttv-cd-bestfit-nresid-std}{1.9}
{n-cd}{23}
{ttv-be-bestfit-resid-min-std}{5}
{ttv-be-bestfit-nresid-std}{1.1}
{ttv-per-c-mean}{8.249164}
{ttv-per-c-std}{0.000003}
{ttv-per-d-mean}{12.401394}
{ttv-per-d-std}{0.000009}
{ttv-per-b-mean}{24.140006}
{ttv-per-b-std}{0.000017}
{ttv-per-e-mean}{48.677714}
{ttv-per-e-std}{0.000053}
{ttv-t0-c-mean}{2231.218}
{ttv-t0-c-std}{0.002}
{ttv-t0-d-mean}{2239.466}
{ttv-t0-d-std}{0.002}
{ttv-t0-b-mean}{2234.093}
{ttv-t0-b-std}{0.002}
{ttv-t0-e-mean}{2263.587}
{ttv-t0-e-std}{0.003}
{ttv-A-cd-mean}{0.066}
{ttv-A-cd-std}{0.001}
{ttv-A-dc-mean}{-0.073}
{ttv-A-dc-std}{0.001}
{ttv-A-be-mean}{0.0491}
{ttv-A-be-std}{0.0005}
{ttv-A-eb-mean}{-0.040}
{ttv-A-eb-std}{0.002}
{ttv-A-cd-min-mean}{96}
{ttv-A-cd-min-std}{2}
{ttv-A-dc-min-mean}{-106}
{ttv-A-dc-min-std}{1}
{ttv-A-be-min-mean}{71}
{ttv-A-be-min-std}{1}
{ttv-A-eb-min-mean}{-58}
{ttv-A-eb-min-std}{3}
{ttv-per-cd-mean}{1604}
{ttv-per-cd-std}{12}
{ttv-per-be-mean}{2852}
{ttv-per-be-std}{50}
{attv-m1-lo}{1.9}
{attv-m1}{2.7}
{attv-m1-err1}{-0.8}
{attv-m1-err2}{1.7}
{attv-r2}{1.18}
{attv-r2-err1}{-0.02}
{attv-r2-err2}{0.02}
{attv-m2}{3.2}
{attv-m2-err1}{-1.0}
{attv-m2-err2}{2.1}
{attv-e1-p95}{0.04}
{attv-e2-p95}{0.03}
{l12ttv-phase}{-0.116}
{l12ttv-phase-err}{0.017}
{l12ttv-r2}{0.8}
{l12ttv-r2-err1}{-0.2}
{l12ttv-r2-err2}{0.4}
{l12ttv-m1e}{31}
{l12ttv-m1e-err1}{-17}
{l12ttv-m1e-err2}{14}
{l12ttv-m2e}{24}
{l12ttv-m2e-err1}{-8}
{l12ttv-m2e-err2}{4}
{l12ttv-logZfreemag}{-3.0}
{l12ttv-logZfreemag-err1}{-0.2}
{l12ttv-logZfreemag-err2}{0.6}
{mass-b-sm22}{203 \pm 60}
{mass-e-sm22}{367 \pm 95}
}[{{\color{red}XX}}]%
}

\begin{document}


\title[A young progenitor system]{A young progenitor for the most common planetary systems in the Galaxy}

\author[1,2,3]{\fnm{John~H.} \sur{Livingston}}\email{john.livingston@nao.ac.jp}

\author[4]{\fnm{Erik~A.} \sur{Petigura}}\email{petigura@astro.ucla.edu}

\author[5]{\fnm{Trevor~J.} \sur{David}}\email{trevorjdavid@gmail.com}

\author[6]{\fnm{Kento} \sur{Masuda}}

\author[7,8]{\fnm{James} \sur{Owen}}

\author[9]{\fnm{David} \sur{Nesvorný}}

\author[10]{\fnm{Konstantin} \sur{Batygin}}

\author[11]{\fnm{Jerome} \sur{de Leon}}

\author[1,2]{\fnm{Mayuko} \sur{Mori}}

\author[11]{\fnm{Kai} \sur{Ikuta}}

\author[11]{\fnm{Akihiko} \sur{Fukui}}

\author[11]{\fnm{Noriharu} \sur{Watanabe}}

\author[12]{\fnm{Jaume~Orell} \sur{Miquel}}

\author[12]{\fnm{Felipe} \sur{Murgas}}

\author[12]{\fnm{Hannu} \sur{Parviainen}}

\author[13]{\fnm{Judith} \sur{Korth}}

\author[12,14]{\fnm{Florence} \sur{Libotte}}

\author[12,14]{\fnm{Néstor~Abreu} \sur{García}}

\author[12,14]{\fnm{Pedro~Pablo~Meni} \sur{Gallardo}}

\author[11]{\fnm{Norio} \sur{Narita}}

\author[12]{\fnm{Enric} \sur{Pallé}}

\author[1,15,2]{\fnm{Motohide} \sur{Tamura}}

\author[16]{\fnm{Atsunori} \sur{Yonehara}}

\author[17]{\fnm{Andrew} \sur{Ridden-Harper}}

\author[18]{\fnm{Allyson} \sur{Bieryla}}

\author[19,20]{\fnm{Alessandro~A.} \sur{Trani}}

\author[21]{\fnm{Eric~E.} \sur{Mamajek}}

\author[22]{\fnm{David~R.} \sur{Ciardi}}

\author[21]{\fnm{Varoujan} \sur{Gorjian}}

\author[10]{\fnm{Lynne~A.} \sur{Hillenbrand}}

\author[22]{\fnm{Luisa~M.} \sur{Rebull}}

\author[23]{\fnm{Elisabeth~R.} \sur{Newton}}

\author[24]{\fnm{Andrew~W.} \sur{Mann}}

\author[25]{\fnm{Andrew} \sur{Vanderburg}}

\author[26]{\fnm{Guðmundur} \sur{Stefánsson}}

\author[27,28]{\fnm{Suvrath} \sur{Mahadevan}}

\author[29]{\fnm{Caleb} \sur{Cañas}}

\author[30]{\fnm{Joe} \sur{Ninan}}

\author[31]{\fnm{Jesus} \sur{Higuera}}

\author[26]{\fnm{Kamen} \sur{Todorov}}

\author[26]{\fnm{Jean-Michel} \sur{Désert}}

\author[32]{\fnm{Lorenzo} \sur{Pino}}

\affil[1]{\orgname{Astrobiology Center}, \orgaddress{\street{2-21-1 Osawa, Mitaka}, \city{Tokyo}, \postcode{181-8588}, \country{Japan}}}

\affil[2]{\orgname{National Astronomical Observatory of Japan}, \orgaddress{\street{2-21-1 Osawa, Mitaka}, \city{Tokyo}, \postcode{181-8588}, \country{Japan}}}

\affil[3]{\orgdiv{Department of Astronomical Science}, \orgname{The Graduate University for Advanced Studies}, \orgaddress{\country{Japan}}}

\affil[4]{\orgdiv{Department of Physics \& Astronomy}, \orgname{University of California}, \orgaddress{\city{Los Angeles}, \state{CA}, \postcode{90095}, \country{USA}}}

\affil[5]{\orgdiv{Center for Computational Astrophysics}, \orgname{Flatiron Institute}, \orgaddress{\street{162 5th Ave}, \city{New York}, \state{NY}, \postcode{10010}, \country{USA}}}

\affil[6]{\orgdiv{Department of Earth and Space Science}, \orgname{Osaka University}, \orgaddress{\city{Osaka}, \postcode{560-0043}, \country{Japan}}}

\affil[7]{\orgdiv{Astrophysics Group}, \orgdiv{Department of Physics}, \orgname{Imperial College London}, \orgaddress{\country{UK}}}

\affil[8]{\orgdiv{Department of Earth, Planetary, and Space Sciences}, \orgname{UCLA}, \orgaddress{\state{CA}, \postcode{90095}, \country{USA}}}

\affil[9]{\orgdiv{Department of Space Studies}, \orgname{Southwest Research Institute}, \orgaddress{\street{1050 Walnut St., Suite 300}, \city{Boulder}, \state{CO}, \postcode{80302}, \country{USA}}}

\affil[10]{\orgdiv{Division of Geological and Planetary Sciences}, \orgname{Caltech}, \orgaddress{\city{Pasadena}, \state{CA}, \postcode{91125}, \country{USA}}}

\affil[11]{\orgdiv{Komaba Institute for Science}, \orgname{University of Tokyo}, \orgaddress{\city{Tokyo}, \postcode{153-8902}, \country{Japan}}}

\affil[12]{\orgname{Instituto de Astrofísica de Canarias}, \orgaddress{\postcode{38200}, \city{La Laguna}, \state{Tenerife}, \country{Spain}}}

\affil[13]{\orgname{Observatoire astronomique de l’Université de Genève}, \orgaddress{\street{Chemin Pegasi 51}, \postcode{1290} \city{Versoix}, \country{Switzerland}}}

\affil[14]{\orgname{Sabadell Astronomical Society}, \orgaddress{\postcode{08206}, \city{Sabadell}, \state{Barcelona}, \country{Spain}}}

\affil[15]{\orgdiv{Department of Astronomy}, \orgname{University of Tokyo}, \orgaddress{\city{Tokyo}, \postcode{113-0033}, \country{Japan}}}

\affil[16]{\orgdiv{Department of Astrophysics and Atmospheric Sciences}, \orgname{Kyoto Sangyo University}, \orgaddress{\street{Kamigamo Motoyama, Kita-ku}, \city{Kyoto}, \postcode{603-8555}, \country{Japan}}}

\affil[17]{\orgname{Las Cumbres Observatory Global Telescope Network}, \orgaddress{\street{6740 Cortona Drive, Suite 102}, \city{Goleta}, \state{CA}, \postcode{93117}, \country{USA}}}

\affil[18]{\orgname{Center for Astrophysics, Harvard \& Smithsonian}, \orgaddress{\street{60 Garden St}, \city{Cambridge}, \state{MA}, \country{USA}}}

\affil[19]{\orgdiv{The Niels Bohr International Academy}, \orgname{Niels Bohr Institute}, \orgaddress{\city{Copenhagen}, \street{Blegdamsvej 17}, \postcode{2100}, \country{Denmark}}}

\affil[20]{\orgname{National Institute for Nuclear Physics}, \orgdiv{Sezione di Trieste}, \orgaddress{\postcode{I-34127}, \city{Trieste}, \country{Italy}}}

\affil[21]{\orgdiv{Jet Propulsion Laboratory}, \orgname{California Institute of Technology}, \orgaddress{\street{4800 Oak Grove Drive}, \city{Pasadena}, \state{CA}, \postcode{91109}, \country{USA}}}

\affil[22]{\orgdiv{NASA Exoplanet Science Institute}, \orgname{IPAC}, \orgaddress{\city{Pasadena}, \state{CA}, \postcode{91125}, \country{USA}}}

\affil[23]{\orgdiv{Department of Physics and Astronomy}, \orgname{Dartmouth College}, \orgaddress{\city{Hanover}, \state{NH}, \country{USA}}}

\affil[24]{\orgdiv{Department of Physics and Astronomy}, \orgname{UNC Chapel Hill}, \orgaddress{\state{NC}, \country{USA}}}

\affil[25]{\orgdiv{Department of Physics}, \orgdiv{Kavli Institute for Astrophysics and Space Science}, \orgname{Massachusetts Institute of Technology}, \orgaddress{\street{77 Massachusetts Ave}, \city{Cambridge}, \state{MA}, \postcode{02139}, \country{USA}}}

\affil[26]{\orgdiv{Anton Pannekoek Institute for Astronomy}, \orgname{University of Amsterdam}, \orgaddress{\street{Science Park 904}, \postcode{1098 XH}, \city{Amsterdam}, \country{The Netherlands}}}

\affil[27]{\orgdiv{Department of Astronomy \& Astrophysics}, \orgname{The Pennsylvania State University}, \orgaddress{\street{525 Davey Laboratory}, \city{University Park}, \postcode{16802}, \state{PA}, \country{USA}}}

\affil[28]{\orgdiv{Center for Exoplanets and Habitable Worlds}, \orgname{The Pennsylvania State University}, \orgaddress{\street{525 Davey Laboratory}, \city{University Park}, \postcode{16802}, \state{PA}, \country{USA}}}

\affil[29]{\orgname{NASA Goddard Space Flight Center}, \orgaddress{\street{8800 Greenbelt Road}, \city{Greenbelt}, \postcode{20771}, \state{MD}, \country{USA}}}

\affil[30]{\orgdiv{Department of Astronomy and Astrophysics}, \orgname{Tata Institute of Fundamental Research}, \orgaddress{\street{Homi Bhabha Road, Colaba}, \city{Mumbai}, \postcode{400005}, \country{India}}}

\affil[31]{\orgname{NSF National Optical-Infrared Astronomy Research Laboratory}, \orgaddress{\street{950 N. Cherry Ave.}, \city{Tucson}, \postcode{85719}, \state{AZ}, \country{USA}}}

\affil[32]{\orgname{INAF—Osservatorio Astrofisico di Arcetri}, \orgaddress{\street{Largo Enrico Fermi 5}, \postcode{I-50125}, \city{Firenze}, \country{Italy}}}


\abstract{
The Galaxy's most common known planetary systems have several Earth-to-Neptune-size planets in compact orbits \citep{Zhu2018}. At small orbital separations, larger planets are less common than their smaller counterparts by an order of magnitude. The young star \sname hosts one such compact planetary system, albeit with four planets that are uncommonly large (5 to 10 Earth radii) \citep{David2019single,David2019multi}. The planets form a chain of near-resonances that result in transit-timing variations of several hours. Here we present a multi-year campaign to characterize this system with transit-timing variations, a method insensitive to the intense magnetic activity of the star. Through targeted observations, we first resolved the previously unknown orbital period of the outermost planet. The full 9-year baseline from these and archival data then enabled robust determination of the masses and orbital parameters for all four planets. We find the planets have low, sub-Neptune masses and nearly circular orbits, implying a dynamically tranquil history. Their low masses and large radii indicate that the inner planets underwent a period of rapid cooling immediately after dispersal of the protoplanetary disk. Still, they are much less dense than mature planets of comparable size. We predict the planets will contract to 1.5--4.0 Earth radii and join the population of super-Earths and sub-Neptunes that nature produces in abundance.
}


\maketitle


\sname is a young (\age), approximately solar-mass ($1.10 \pm 0.05$~M$_\odot$) star in the Taurus star-forming region \citep{Wichmann1996, David2019single, Gaidos2022, Johnson2022, SuarezMascareno2022, Maggio2022}. Observations by NASA’s \kepler space telescope in its extended \ktwo mission \citep{Howell2014} revealed transits of the star by four different planets, each larger than Neptune \citep{David2019single, David2019multi}. The \sname planets occupy a sparsely populated region of the observed exoplanet period versus radius plane. As a young system of large planets, it provides a crucial snapshot of planetary architecture just after formation, serving as the `missing link' between protoplanetary disks and the mature systems found by \kepler \citep{David2019multi}. Measuring their masses and orbits is, therefore, a key test of planet formation theories and allows us to witness early evolutionary processes, such as atmospheric mass loss, that sculpt planetary systems over billion year timescales.

\begin{figure}[H]
\centering
\includegraphics[width=\linewidth]{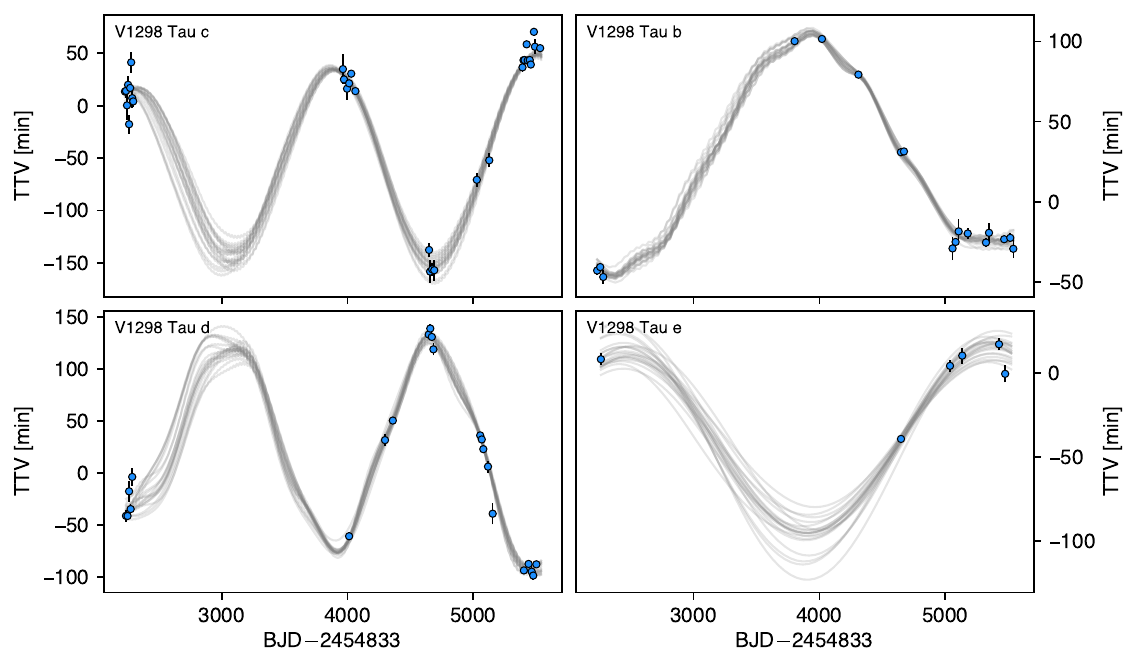}
\caption{\textbf{Transit timing variations in the \sname system.} Top left, points show the transit times of planet c measured against a reference linear ephemeris; error bars represent 1$\sigma$ uncertainties. Grey curves show credible transit times drawn from the $N$-body models described in the text. Bottom left, same as above but for planet d. The interactions between c and d are nearly sinusoidal and anticorrelated. Top and bottom right, the same but for planets b and e. The TTVs of b and e are also sinusoidal and anticorrelated.}
\label{fig:1}
\end{figure}

Between 2019 and 2024, we observed 43 other transits of all four planets using both space- and ground-based telescopes. This campaign successfully recovered the previously lost outermost planet, \sname~e, resolving a long-standing period ambiguity \citep{Feinstein2022} (Methods). We performed a homogeneous and self-consistent modelling of all transit data from 2015 to 2024. After determining the transit shape parameters, we fit the midpoint of each transit individually (Methods). The transit-timing variations (TTVs) are shown in Figure~\ref{fig:1}. All planets exhibit significant TTVs with amplitudes ranging from approximately 50 to 100~minutes. Moreover, the TTVs of the c--d pair are anticorrelated, as are those of the b--e pair. This indicates that the c--d and b--e interactions dominate over other pairwise interactions. 

Previous works have developed analytic models of TTVs applicable to certain orbital configurations \cite{Lithwick2012,Nesvorny2016}. Other works have developed $N$-body TTV models based on a dynamical integration of the star-planet system subject to Newtonian gravity \citep[e.g.][]{Carter2012,Deck2014,Mills2016}. Analytic models are generally faster to evaluate, have fewer free parameters, and offer a clearer connection between the system properties and the TTV waveform. $N$-body models, with more free parameters, are slower to evaluate but can completely describe any star--planet system.

Most TTV studies in the literature treat planets from the four-year \kepler mission. These studies had the benefit of near-continuous sampling over a full TTV period. The sparse sampling of our dataset presents different challenges. We, therefore, used analytic models to build our intuition of the system's dynamics before undertaking a full $N$-body analysis. The nature of TTV interactions depends on the proximity to resonance $\Delta$, defined as $\Delta = \frac{P_2}{P_1}\frac{j - 1}{j} - 1$, where $P_1$ and $P_2$ are the orbital periods of the inner and outer planets, respectively, and $j$ is a positive integer defining the resonance, with smaller $\Delta$ associated with larger TTVs. In this system,
$\Delta_\mathrm{cd} = 0.2$\%, 
$\Delta_\mathrm{db} = -2.7$\%, 
and $\Delta_\mathrm{be} = 0.8$\%.
Given the strength of the c--d and b--e interactions over d--b interactions, we used analytic models to treat both the c--d and b--e interactions separately. 
The analytic models indicated low masses and low eccentricities (Methods).

Guided by our analytic results, we then performed a full $N$-body dynamical fit to the transit times to derive a final, robust set of planet parameters. This model accounts for all gravitational interactions in the system simultaneously, including subtle, higher-order TTVs that can help break degeneracies inherent in our analytic models. Details of our $N$-body model, Bayesian statistical framework, and Markov chain Monte Carlo sampling are provided in Methods.

Figure~\ref{fig:1} shows a selection of credible models drawn from our posterior samples, along with the timing data. The models fit the data well, with increased scatter where observations are sparse. The credible range of each planetary parameter is listed in Table~\ref{tab:1}. We find masses of 
$M_c = \val{mass-c}$~\Me, 
$M_d = \val{mass-d}$~\Me,
$M_b = \val{mass-b}$~\Me, and
$M_e = \val{mass-e}$~\Me
(the uncertainties correspond to the $68\%$ highest density intervals of the marginal posteriors). In addition, the planetary eccentricities are all less than about 1\%. A detailed dynamical analysis (see Methods) confirms that this solution corresponds to a long-term stable and non-resonant orbital architecture. The $N$-body results are consistent with the analytic results at $2\sigma$ or better with smaller uncertainties. The $N$-body model is a more complete description of the planetary dynamics than our analytic models and includes effects like synodic chopping \citep{Deck2015}, as well as d-b interactions. It exhibits root mean square (r.m.s.) values of 4--11 min, consistent with our analytic models. Henceforth, we adopt and interpret the $N$-body results. The masses of planets b and e have substantial fractional uncertainties but are distinct from zero; they are larger than 4.8~\Me and 7.8~\Me to 95\% confidence. The broad uncertainties stem from the well-known mass versus eccentricity degeneracy \citep{Lithwick2012}. Extremely low masses and high eccentricities would produce b-d interactions that are inconsistent with the data. 

\begin{table}[ht!]
\footnotesize
\begin{center}
\begin{minipage}{\textwidth}
\caption{\textbf{Planetary parameters}}\label{tab:1}
\begin{tabular}{llcccc}
\toprule
Parameter & Unit & c & d & b & e \\
\midrule
$M_p$ & \Me & $\val{mass-c}$ & $\val{mass-d}$ & $\val{mass-b}$ & $\val{mass-e}$ \\
$R_p$ & $R_\oplus$ & $\val{radius-c}$ & $\val{radius-d}$ & $\val{radius-b}$ & $\val{radius-e}$ \\
$\rho$ & \gcc & $\val{rho-c}$ & $\val{rho-d}$ & $\val{rho-b}$ & $\val{rho-e}$ \\
$e$ & \% & $<0.94$ & $<0.87$ & $0.79 \pm 0.41$ & $<1.24$ \\
$P$ & days & \val{ttv-per-c-mean}(3) & \val{ttv-per-d-mean}(9) & \val{ttv-per-b-mean}(17) & \val{ttv-per-e-mean}(53) \\
$a$ & AU & $0.0824 \pm 0.0012$ & $0.1081 \pm 0.0016$ & $0.1685 \pm 0.0025$ & $0.2689 \pm 0.0040$ \\
$T_{\mathrm{eq}}$ & K & $953 \pm 36$ & $831 \pm 31$ & $666 \pm 25$ & $527 \pm 20$ \\
\botrule
\end{tabular}
\end{minipage}
\end{center}
\footnotetext{Parameter definitions: $M_p$, planetary mass; $R_p$, planetary radius; $\rho$, bulk density; $e$, orbital eccentricity; $P$, orbital period; $a$, semi-major axis; $T_{\mathrm{eq}}$, equilibrium temperature; and $T_{\mathrm{eff}}$, stellar effective temperature. Uncertainties denote the 68\% credible interval and upper limits are $1\sigma$. $T_{\mathrm{eq}}$ values assume $T_\mathrm{eff}=4970\pm120\,K$ \citep{David2019single} and zero Bond albedo. Planet-to-star mass and radius ratios are converted to planetary masses and radii using $M_\star = 1.10\pm0.05\,M_\odot$ and $R_\star = 1.32\pm0.05\,R_\odot$ \citep{David2019single}. Eccentricities are given in percent.}
\end{table}

The combination of low planet masses with the youth of the host star makes Doppler mass measurements challenging. The expected semi-amplitudes of the radial velocity are approximately 1--2~\ms, which are two orders of magnitude smaller than the stellar activity signal. For comparison, the measured r.m.s. of the radial velocity in existing data sets for \sname is 260 \ms, 197 \ms, and 195 \ms for HARPS-N, CARMENES VIS, and CARMENES NIR, respectively \citep{SuarezMascareno2022}. The challenge of this high stellar activity has been a central theme in recent studies of the system \citep{Sikora2023, Finociety2023, DiMaio2024, Biagini2024}. \cite{SuarezMascareno2022} simultaneously modelled planetary and activity radial velocity variations, reporting masses of $\val{mass-b-sm22}$~\Me and $\val{mass-e-sm22}$~\Me for planets b and e, an order of magnitude larger than our TTV results. However, \cite{Blunt2023} found that the planet-activity model is biased toward over-predicting planet masses when stellar activity dominates. Given these challenges and the risk of systematic bias, a TTV-only analysis provides, at present, the most robust and unbiased mass constraints for this system. Notably, our dynamical mass for planet b is consistent with independent atmospheric constraints. A recent analysis of transmission spectra captured by the James Webb Space Telescope \cite{Barat2025} inferred a mass from the atmospheric scale height that is in excellent agreement with our TTV result. That two independent methods---one based on gravitational dynamics and the other on atmospheric structure---yield such consistent results provides a powerful validation of our measurement.

\begin{figure}[H]
\centering
\includegraphics[width=\linewidth]{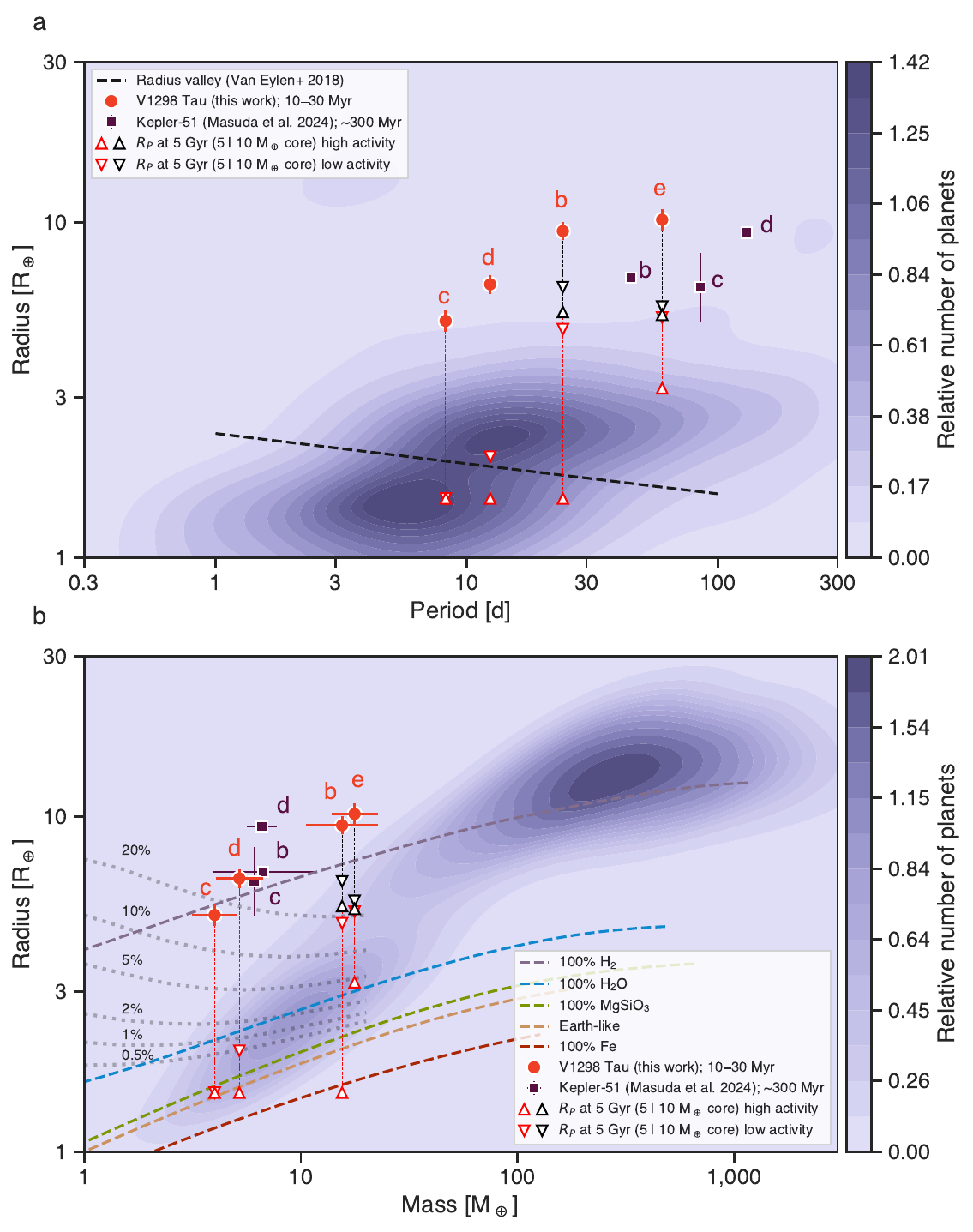}
\caption{\textbf{Planetary radius versus orbital period and planetary mass.} \textbf{a,b,} Planetary radius versus orbital period (\textbf{a}) and planetary radius versus planetary mass (\textbf{b}) for the \sname system (red filled circles); error bars represent 1$\sigma$ uncertainties. The low-density planets of the Kepler-51 system are shown for comparison (purple squares), along with kernel density estimates of the distributions of well-characterized exoplanets (shaded contours), drawn from the NASA Exoplanet Archive (n=624 planets with mass and radius uncertainties less than 20\%, P\,$<$\,150 days, and host \teff\,=\,4500–6500 K to exclude M dwarfs). The parameters of the Kepler-51 planets were sourced from the `outside 2:1' solution in Table 6 of \cite{Masuda2024}. Theoretical radius evolution tracks from \cite{Poppenhaeger2021} are shown as vertical dashed lines. The terminal radii at 5~Gyr from that work are shown as open triangles. The colour indicates the assumed core mass (red for 5~\Me and black for 10~\Me). The orientation represents the stellar extreme-ultraviolet activity level (upwards for high activity, downwards for low activity). The black dashed line in \textbf{a} depicts the observed location of the radius valley \citep{vanEylen2018}. Theoretical mass–radius relations for different planet compositions from \cite{Zeng2019} are shown in \textbf{b} as dashed lines. Grey dotted lines indicate theoretical mass-radius relations for Earth-like cores with H/He envelopes with various mass fractions from \cite{LopezFortney2014}, calculated for an age of 100~Myr and an insolation of 10~F$_\oplus$.}
\label{fig:context}
\end{figure}

The planetary densities that we measured in the \sname system are among the lowest exoplanet densities recorded. The only known multi-planet system exhibiting comparably low densities is, perhaps not coincidentally, the young (approximately 300~Myr) transiting system Kepler-51, which also permitted mass measurements through TTVs \citep{Steffen2013, Masuda2014, Hadden2017, LibbyRoberts2020, Masuda2024}, although \sname is significantly younger and more compact. Figure~\ref{fig:context} places the \sname planetary system in the context of the broader, mature exoplanet population. Figure~\ref{fig:context}a shows these young planets positioned above the radius gap \citep{vanEylen2018}. To trace their future evolution, we overplot the `fluffy' planet scenarios from \cite{Poppenhaeger2021}, which are the most relevant analogues. These models bracket a range of possibilities by assuming two different core masses (5~\Me and 10~\Me) and two stellar extreme-ultraviolet activity levels that result in different degrees of atmospheric mass loss. The 5~\Me scenario is a particularly strong analogue, as our own interior structure modelling (see Methods) constrains the core masses of planets c and d to be 4--6~\Me (1-$\sigma$). The resulting tracks suggest that some planets will contract across the gap to become super-Earths, while others will become sub-Neptunes---thus directly tracing the formation of the bimodal radius distribution observed by \kepler. Figure~\ref{fig:context}b reveals substantial H/He envelopes \citep{LopezFortney2014}, though their final evolved states may have densities that are degenerate with water worlds \citep{Zeng2019}.

The low masses and densities of the \sname planets have significant ramifications for planet formation theory. Theoretical modelling indicated planet c ($M_c = \val{mass-c}$~\Me) was one of the best targets for constraining its formation history: a mass higher than 10~\Me would be consistent with standard core-accretion models, whereas a mass lower than 6~\Me would require a ‘boil-off’ phase during protoplanetary disk dispersal \cite{Owen2020}. Such a phase occurs when the pressure support of the disk is removed swiftly, triggering profuse atmospheric mass loss through a Parker wind and rapid cooling, leaving behind an envelope with lower entropy and a longer Kelvin-Helmholtz timescale compared with predictions from standard core-accretion models \citep{OwenWu2016,Rogers2024}.

To explore the possible formation channels for the \sname planets, we modelled the planets as two-layer objects consisting of an Earth composition rocky core ensheathed in a H/He envelope. The initial envelope entropy is parameterized by its Kelvin-Helmholtz contraction timescale. We ran a dense grid of models spanning core mass, initial envelope mass fraction, and initial envelope entropy at the location of each planet in the system and evolved them to the current age of the system. 

\begin{figure}[H]
\centering
\includegraphics[width=\linewidth]{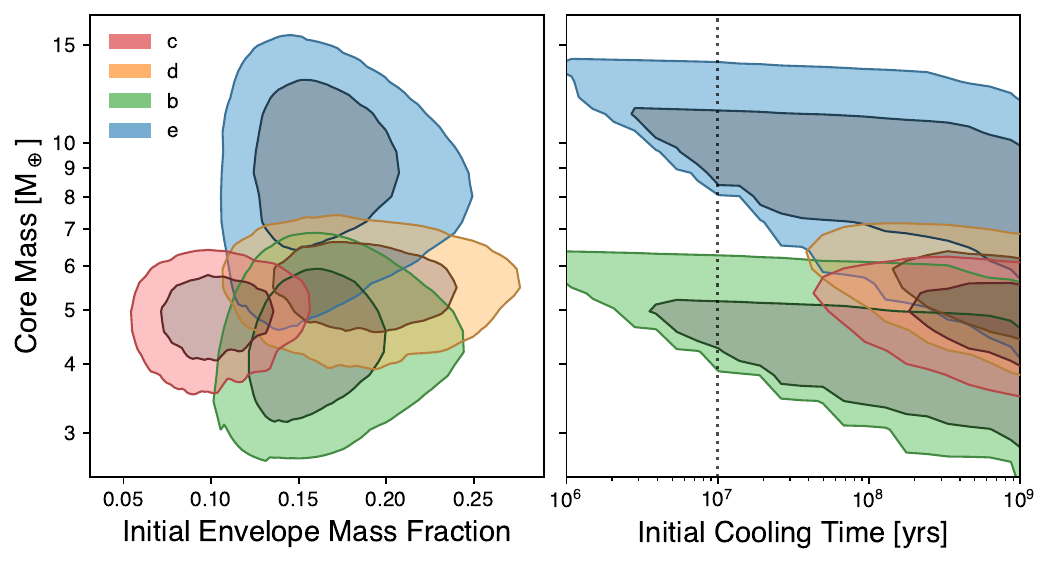}
\caption{\textbf{Posterior distributions for the initial properties of the \sname planets.} The posteriors are derived by applying the planetary evolution and mass loss framework of \cite{Owen2020} to our measured masses and radii for planets c (red), d (orange), b (green), and e (blue). Left, initial envelope mass fraction versus core mass. Right, initial Kelvin-Helmholtz cooling timescale versus core mass. Contours show the 1$\sigma$ and 2$\sigma$ credible regions. (Note that the jagged appearance of some contours is a numerical artefact of the discrete core mass grid used in our analysis; see Methods for more details.) The vertical dotted line in the right panel at 10 Myr marks the approximate upper limit for standard high-entropy formation models. These models are strongly disfavoured for the inner planets c and d, whereas for the less-irradiated outer planets b and e, the method lacks the statistical power to distinguish between high- and low-entropy scenarios.}
\label{fig:evolution}
\end{figure}

Figure~\ref{fig:evolution} shows posterior distributions for the initial properties of all four planets, providing a deeper insight into the system architecture. The right panel confirms that the inner planets c and d require low-entropy initial states (much greater than 30~Myr Kelvin-Helmholtz cooling times), whereas the less-irradiated outer planets b and e remain unconstrained. The left panel, however, reveals a notable uniformity: all four planets are consistent with having similar core masses (approximately 4--6~\Me) and initial envelope mass fractions (approximately 0.1--0.2). This indicates the system is an exemplar of the `peas in a pod' phenomenon at formation \citep{Weiss2018}, implying its present-day size diversity is a transitory phase driven by different levels of photoevaporation.

Extended Data Fig.~\ref{edfig:1} shows that the measured masses and radii of planets c and d lie outside the region of parameter space accessible to standard, high-entropy core-accretion models. As the illustrative tracks in the figure demonstrate, only lower-entropy (boil-off) models can simultaneously satisfy both the mass and radius constraints after accounting for 23~Myr of evolution and mass loss. During boil-off, the planetary envelope becomes over-pressurized and expands hydrodynamically, carrying away significant internal energy and leaving behind a cooler interior \citep{OwenWu2016,Rogers2024,Tang2024}. Although our measurements support boil-off for the inner planets, recent atmospheric retrievals indicating a high internal temperature for planet b \citep{Barat2024a} present a possible tension that merits further investigation.

Theoretical modelling of the system under the influence of extreme ultraviolet- and X-ray-driven photoevaporation indicates these planets will continue to lose mass over the next 100 million years \citep{Poppenhaeger2021, Maggio2022}, even though they have already experienced significant atmospheric loss. For our measured masses, standard evolutionary models predict that all the planets will retain a small fraction of their initial atmospheres, although the inner two could become stripped, depending on the future spin evolution of the star \citep{Poppenhaeger2021}. Interestingly, observational searches for continuing atmospheric escape have so far yielded inconclusive results \citep{Vissapragada2021, Feinstein2021, Gaidos2022, Alam2024}, possibly because strong stellar winds act to suppress planetary outflows \citep{Vidotto2020, Carolan2020, WangDai2021}.


\clearpage


\section*{Methods} \label{sec:methods}

\subsection*{Transit observations and analysis}

We analysed a heterogeneous dataset of light curves from space- and ground-based telescopes (see Supplementary Table~\ref{sitab:1}) to measure transit times for the ultimate purpose of modelling TTVs. We used \texttt{PyMC3} \citep{pymc3}, \texttt{exoplanet}\footnote{\url{https://docs.exoplanet.codes/en/stable/}} \citep{exoplanet_joss}, and \texttt{starry} \citep{luger2019} for light curve fitting, incorporating tailored models for correlated noise and instrumental systematics appropriate for each dataset.

Our analysis of the \ktwo and \tess light curves involved two distinct approaches with different noise models. For the joint analysis of all transits in both light curves (described below), we modelled stellar variability as a Gaussian Process \citep[GP;][]{Rasmussen2005}. In contrast, for measuring individual transit times (see below), a third-order basis spline was sufficient to model the local correlated noise.

To account for systematics in the \spitzer data we used the pixel-level decorrelation method \citep[PLD;][]{Deming2015}, which amounts to a linear model with a design matrix formed by the PLD basis vectors (see below for more details). For the ground-based datasets we included a linear model with a design matrix formed by airmass, pixel centroids, and the pixel response function peak and width covariates, when available. 

Limb darkening coefficients were calculated using stellar parameters from \cite{David2019single} by interpolation of the parameters tabulated by \cite{Claret2012,Claret2017}. These were fixed for individual transit fits but sampled with uninformative priors in the joint \ktwo-\tess analysis described below.

We used {\tt BFGS} optimization \citep{NoceWright2006} as implemented in {\tt scipy.optimize} for initial parameter estimates, followed by posterior sampling with \citep[NUTS,][]{HoffmanGelman2014}, an efficient gradient-based Hamiltonian Monte Carlo (HMC) sampler implemented in {\tt PyMC3}. Chains were well-mixed (Gelman-Rubin statistic less than about 1.01) with negligible sampling error.

We first performed a joint fit of \ktwo and \tess data assuming a linear ephemeris (see below). We then measured all individual transit times uniformly using Gaussian priors from the joint fit for $R_p/R_\star$, $b$ (the transit impact parameter), and $T_{14}$ (the total transit duration) and uniform priors for $T_c$ (the transit centre time) centred on predicted times. We verified that $T_c$ posteriors were Gaussian and well-isolated from prior edges. 

In all individual transit fits, we assumed Gaussian iid noise and included a jitter parameter \isr{\sigma}{jit} to account for underestimated photometric uncertainties. The log-likelihood was thus:
\begin{equation*}
    \ln{\cal{L}} = -\frac{1}{2}\ln{\lvert\boldsymbol{\Sigma}\rvert} - \frac{1}{2}\boldsymbol{r}^T\boldsymbol{\Sigma}^{-1}\boldsymbol{r} + \mathrm{const.},
\end{equation*}
where $\boldsymbol{\Sigma}$ is the diagonal covariance matrix with entries equal to the total variance (that is, the $i$th entry is $\isr{\sigma}{tot,i}^2$ = $\isr{\sigma}{obs,i}^2+\isr{\sigma}{jit}^2$, where $\isr{\sigma}{obs,i}$ is the observational uncertainty of the $i$th data point), and $\boldsymbol{r}$ is the residual vector ($\boldsymbol{r} = \lbrack \hat{y}_1 - y_1, \hat{y}_2 - y_2, ..., \hat{y}_n - y_n \rbrack$, where $\hat{y}$ is the model and $y$ is the data consisting of $n$ measurements). When a given transit event was observed by multiple telescopes (e.g., from the Las Cumbres Observatory, LCO, \cite{Brown2013}) or multiple bandpasses from the same instrument (e.g., from MuSCAT3, \cite{Narita2020})), we jointly fit all light curves covering the same event.

We obtained ground-based follow-up transit observations from a variety of facilities spanning multiple observing seasons. Early in the project, observations were distributed diversely among a half-dozen telescopes, but later on we focused almost exclusively on the LCO telescope network, which enabled both the acquisition of data and its analysis to be conducted more uniformly. The individual dates, facilities, bandpasses, and exposure times of these observations are listed in chronological order in Supplementary Table~\ref{sitab:1}.
The measured transit times are provided in Supplementary Table~\ref{sitab:2}.

\subsubsection*{Joint analysis of the \ktwo and \tess light curves}
\label{sec:k2-tess}

\sname (EPIC~210818897) was observed between 7 February 2015 and 23 April 2015 during Campaign 4 of the \ktwo mission \citep{Howell2014}. We analysed the \ktwo light curve produced by the \texttt{EVEREST}\footnote{\url{https://github.com/rodluger/everest}} pipeline \citep{Luger2016,Luger2018}, which is available at the Mikulski Archive for Space Telescopes (MAST).\footnote{\url{https://archive.stsci.edu/hlsp/everest}}. 

\sname (TIC 15756231) was observed at 2-minute cadence in Sectors 43-44 (16 Sep - 6 Nov 2021) of the \tess mission \citep{Ricker2015} as part of the Director’s Discretionary Time (DDT) programme \#036 (P.I. T. David). 

We conducted a joint fit to the \ktwo and \tess light curves assuming a linear ephemeris, using a GP to account for correlated noise arising from a combination of stellar variability and instrumental systematics.\footnote{Based on the tutorial available at \url{https://gallery.exoplanet.codes/tutorials/lc-multi/}} 
We used a simple harmonic oscillator (SHO) covariance function with power spectral density given by:
\begin{equation*}
    S(\omega) = \sqrt{\frac{2}{\pi}} \frac{S_0\,\omega_0^4}
    {(\omega^2-{\omega_0}^2)^2 + {\omega_0}^2\,\omega^2/Q^2},
\end{equation*}
where $\omega$ is the angular frequency, $\omega_0$ is the undamped angular frequency of the oscillator, and $S_0$ is a scale factor that sets the amplitude of the variability. This was re-parameterized by the undamped period of the oscillator $\rho$ (defined as $\rho = 2\,\pi / \omega_0$), the standard deviation of the process $\sigma$ (defined as $\sigma = \sqrt{S_0\,\omega_0\,Q}$), and the quality factor $Q$ (fixed to $\frac{1}{3}$). Similar to our model for individual transits, we included a photometric jitter term (\isr{\sigma}{jit}), the square of which was added to the diagonal of the covariance matrix. The likelihood was thus identical to that shown for individual transits above, but the covariance matrix contained non-zero off-diagonal elements determined by the covariance function. The results of this fit are shown in Extended Data Fig.~\ref{edfig:2} and the posteriors are summarized in Extended Data Table~\ref{edtab:1}.

\subsubsection*{Individual \ktwo and \tess transits}
\label{sec:ktwo-tess-individual}

To create a uniform transit timing dataset, we analysed individual transits from the long-baseline \ktwo and \tess light curves in the same manner as our short-duration follow-up observations. We constructed individual datasets from windows of three times the transit duration centred on each transit event. In the case of overlapping transits, we used the longest transit duration and centred the window on the approximate midpoint of the dimming event.
Unlike follow-up datasets, which are often partial transits, stellar variability is typically nonlinear on the timescale of these datasets. To account for this, we included a third-order basis spline with 5 evenly-spaced knots.
The transits of \sname~c on 10 and 26 October 2021 UT resulted in poor quality fits, likely due to the presence of short-timescale red noise close to ingress/egress and/or low SNR; as the timing posteriors from these fits were highly non-Gaussian, we discarded them from subsequent analyses.

\subsubsection*{\spitzer}
\label{sec:spitzer}

We used the ephemeris derived from the \ktwo observations \citep{David2019single} to predict transits of \sname~b within \spitzer visibility windows in 2019. Subsequently, we did the same for \sname~c,d using the ephemerides from \cite{David2019multi}. An additional transit of \sname~b was scheduled in early 2020 using an updated ephemeris based on \cite{David2019single} and the first \spitzer observation of that planet. The \spitzer data and best-fit transit models are shown in Extended Data Fig.~\ref{edfig:3}.

The first epoch of \spitzer observations of \sname were acquired as part of the DDT programme 14227 (P.I. E. Mamajek) and executed on 1 June 2019 UT. The second epoch of \spitzer observations were acquired as part of the ToO Program 14011 (P.I. E. Newton) and executed on 28 December 2019 UT. In both epochs, data were acquired with channel 2 of the infrared array camera (IRAC) onboard Spitzer (with effective wavelength $\lambda_\mathrm{eff} = 4.5$~\micron) in the subarray mode using 2-s exposures. A third epoch of \spitzer observations were acquired in IRAC channel 1 ($\lambda_\mathrm{eff} = 3.6$~\micron) as part of DDT 14276 (P.I. K. Todorov) and executed on 4 January 2020 UT.

We extracted photometry following \cite{Livingston2019b} and modelled instrumental systematics using PLD, which combines normalized pixel light curves as basis vectors in a linear model:
\begin{align*}
    M_\mathrm{PLD}^t(\boldsymbol{\alpha}) = \frac{\sum_{i=1}^9 c_i P_i^t}{\sum_{i=1}^9 P_i^t},
\end{align*}
where $P_i$ is the light curve of the $i$th pixel, the superscript $t$ denotes the value at a specific time step, and $\boldsymbol{\alpha} = \{c_1, ..., c_9\}$ are the coefficients of the PLD basis vectors.
The first epoch, which captured a partial transit of \sname~b over approximately 11.5 hours, was well-fit by including a linear trend in addition to PLD. The second epoch, which contained transits of both planets c and d over a approximately 14-hour baseline, exhibited significant nonlinear variability that required the inclusion of a basis spline. Similarly, the third epoch, containing a full transit of planet b over approximately 12.5 hours, also warranted the inclusion of a basis spline; although IRAC1 systematics are typically larger than those of IRAC2, PLD performed well and we attribute this to stellar variability. We validated our approach of selecting the baseline model by inspecting the fit residuals for the longest and most complex observation (the second epoch), where a quantitative comparison confirmed that a basis spline was strongly preferred over a simple linear trend by the Bayesian Information Criterion \citep{Schwarz1978}.

\subsubsection*{Ground-based observations}
\label{sec:ground}

The majority of our follow-up transit observations were obtained from 2020 to 2024 using LCO. We primarily used the Sinistro \cite{Brown2013} and MuSCAT3 instruments on the 1m and 2m telescopes, respectively. 

In addition to LCO, we utilized data from a variety of other facilities, including: APO/ARCTIC \citep{Huehnerhoff2016}; FLWO/KeplerCam \citep{gaborthesis}; WIYN/HDI \citep{hdi2013}; TMMT \cite{monson2017}; MuSCAT \citep{Narita2015}; MuSCAT2 \citep{Narita2019}; Araki/ADLER.
Data were obtained using a variety of filters and reduced using standard pipelines and methods \citep{Fukui2011,stefansson2017,stefansson2018,Collins2017,McCully2018,Stefansson2020}. 
See Supplementary Information for more details.

\subsubsection*{Recovering Planet e}
\label{sec:planet-e}

The outermost planet, \sname~e, transited only once during the K2 mission. \tess recovered transits of all four planets, including a second transit of planet e \citep{Feinstein2022}. 
It was not clear how many transits occurred between the \ktwo and \tess observations given the 6.5 year gap between the two campaigns. 
Thus, a discrete comb of periods was allowed, such that $P$ = $\Delta t / n$, where $\Delta t$ is the measured time between transit midpoints and the integer $n$ = 1, 2, 3, ..., \textbf{$n_\mathrm{max}$}. The upper bound on $n$, and thus lower bound on period of 42.7 days, is provided by the absence of additional transits by planet e within the \ktwo and \tess time series \citep{Feinstein2022}.

By summer of 2022, a preliminary version of our timing dataset had revealed large TTVs of planet b that we assumed were dominated by interactions with planet e. We ran a suite of TTV models at each of the possible $P_e$ between 42.7 and 120 days. Few trial periods yielded good fits to the timing dataset, and dynamical simulations revealed only a fraction of those were stable over $\mathcal{O}(10^6)$ years. One of the stable solutions with $P_e = 48.7$~days corresponded to a near 2:1 commensurability for the b-e pair, a common configuration among the \kepler planets exhibiting large/detectable TTVs. With this prediction, we recovered a partial transit of planet e from the ground on October 18, 2022. LCO datasets used to recover planet e and confirm its orbital period are shown in Extended Data Fig.~\ref{edfig:4}.

\subsubsection*{Datasets containing flares}
\label{sec:flares}

Several observations were affected by stellar flares and were excluded from our TTV analyses to avoid potential timing measurement biases. These datasets were modelled using our standard approach, augmented with a parametric flare model \citep{Davenport2014} (see Extended Data Fig.~\ref{edfig:5}).
Significant flares were observed in ARCTIC data (2020 October 12; see also \cite{Vissapragada2021}), KeplerCam data (2023 September 24), and LCO data (2023 December 18), with amplitudes ranging from 6 parts per thousand (ppt) to 42 ppt and timescales of 14 min to 21 min. The parameters of these flares are detailed in the Supplementary Information, and may prove valuable for future studies of the activity of \sname. 

\subsection*{Mass constraints from analytic TTV modelling}
\label{sec:multi-harmonic}

To build intuition for the system's dynamics, we first performed a preliminary analysis using analytic models of transit-timing variations (TTVs). Based on the foundational analytic frameworks for TTVs \citep{Agol05,Lithwick2012,Nesvorny2016}, we determined that the system's dynamics can be effectively decoupled into two pairs of planets: c-d and b-e.

To quantify the TTV behaviour, we fitted a multi-harmonic sinusoidal model to the transit timeseries (see Supplementary Information for model equations). We explored the posterior distributions of the 16 model parameters using a Markov Chain Monte Carlo (MCMC) sampler, similar to approaches used by other public TTV analysis codes \citep{Deck2016,Agol2021}. The posteriors of these parameters are listed in Supplementary Table~\ref{sitab:3}, and the model fits are shown in Extended Data Fig.~\ref{edfig:6}.

The results for the c-d pair are consistent with the planets being in a near-resonant regime. The TTVs are well-described by a single sinusoid with a period of $\isr{P}{cd} = \val{ttv-per-cd-mean}\pm \val{ttv-per-cd-std}$~days and an RMS of the residuals of only $\val{ttv-cd-bestfit-resid-min-std}$~min. This sinusoidal signal is dominated by variations in the planets' mean longitudes ($\lambda$), characteristic of systems very close to resonance. The ratio of the TTV amplitudes is sensitive to the planetary mass ratio, indicating nearly equal masses ($M_d/M_c \approx 1.2$). From the full fit, we derived preliminary masses of $M_c \approx \val{attv-m1}^{+\val{attv-m1-err2}}_{\val{attv-m1-err1}}$~\Me and $M_d \approx \val{attv-m2}^{+\val{attv-m2-err2}}_{\val{attv-m2-err1}}$~\Me.

In contrast, the b-e pair is well-described by a simpler, linear TTV model, as it is further from resonance. The TTVs arising from variations in mean longitude and eccentricity have the same frequency in this regime. In this case, a well-known degeneracy exists between the planet masses and their orbital eccentricities \citep{Deck2015,Hadden2016}, leading to broader initial constraints of $M_b = \val{l12ttv-m1e}_{\val{l12ttv-m1e-err1}}^{+\val{l12ttv-m1e-err2}}$~\Me and $M_e = \val{l12ttv-m2e}_{\val{l12ttv-m2e-err1}}^{+\val{l12ttv-m2e-err2}}$~\Me. A full theoretical treatment as well as a discussion of strategies to break the mass-eccentricity degeneracy, such as measuring secondary eclipse times \citep{Winn2010}, can be found in the Supplementary Information.

\subsection*{Mass constraints from $N$-body TTV modelling}

Guided by our analytic models, our primary analysis relies on a full N-body dynamical model to derive the final planet parameters. We fit the model to the observed transit times using a Bayesian framework. To be robust against outlier measurements, we adopted a log-likelihood function based on the Student's t-distribution \citep{2016ApJ...820...39J, 2021PSJ.....2....1A}, with priors as listed in Extended Data Table~\ref{edtab:2}. The posterior probability distribution was sampled using a No-U-Turn Sampler \citep{DUANE1987216, 2017arXiv170102434B}. The model is implemented in \texttt{JAX} to enable automatic differentiation and is available as part of the \texttt{jnkepler} package \citep{jnkepler,Masuda2024}. The full mathematical details of the model implementation, the log-likelihood equation, and the sampler setup are provided in the Supplementary Information. The resulting mass and eccentricity posterior distributions are shown in Extended Data Fig.~\ref{edfig:7}.

To verify the physical plausibility of our solution, we performed a detailed dynamical analysis of the posterior. We investigated both the long-term stability and the resonant state of the system using multiple complementary methods.
First, to assess stability, we used a probabilistic classifier \citep[\texttt{SPOCK;}][]{Tamayo2020} on 1000 samples from our posterior, which yielded a median stability probability of 95\% over 10$^9$ orbits. We confirmed this with direct N-body integrations of 128 samples for 1~Myr, which showed the system is deeply stable and regular (minimum separation over 12~R$_H$ (mutual Hill radii); maximum semimajor axis drift of less than 0.01\%; \texttt{MEGNO} (Mean Exponential Growth factor of Nearby Orbits) = 2.000). As a final check, we integrated 32 posterior samples for 4 Myr, all of which were found to be stable.
Second, to characterize the resonant state, our integrations show that all classical resonant angles are circulating, which we confirmed by projecting our solution onto the resonant representative plane \citep{Nesvorny2016}. The solution lies clearly outside the resonant island where libration would occur, confirming the non-resonant nature of the system.

\subsection*{Initial thermal state and planetary evolution}

Young planets with hydrogen dominated atmospheres contract over time due to mass-loss and thermal evolution. \cite{Owen2020} showed that young planets with measured masses and radii can be used to constrain their initial entropies. Planets with a measured mass and radius have a degeneracy between their hydrogen envelope mass fraction and their thermal state, where the planet's hydrogen envelope mass can be reduced and compensated for by an increase in its entropy. However, this can only go so far; the envelope mass cannot be reduced arbitrarily to the point where it is too small to survive mass-loss. Thus, one can place a bound on the initial entropy of the planet such that it survives until today. To perform this calculation, we compute a grid of {\sc mesa} evolutionary models including photoevaporation (comparison to \cite{Owen2024} indicates that these planets will be undergoing photoevaporation rather than core-powered mass loss). This model grid comprised 36 core-masses, 128 initial mass fractions, and 96 initial entropies. We use an identical method to \cite{Owen2020}. We then compared this model grid to the observed mass and radii of the \sname planets to derive posterior distributions of the core-masses, initial envelope mass fractions, and initial entropies (which we encode as the initial Kelvin-Helmholtz cooling timescale of the planets). Our results indicate all the planets have initial envelope mass fractions and core masses that are consistent with typical sub-Neptunes at billion year ages. Furthermore, the initial cooling timescales are constrained to require boil-off for planet's c \& d, whereas an evolution without boil-off cannot be ruled out for the outer planets. Extended Data Fig.~\ref{edfig:8} shows the models that best reproduce the present-day mass and radius of planets c \& d. Interestingly, they require initial low entropy; that is, an initial Kelvin-Helmholtz contraction time that is longer than the age of the system. Furthermore, if one only considers models with high initial entropy, one can match the current mass or radius, but not both.

\newpage


\backmatter
 
\bmhead{Data Availability}
The transit times that form the primary dataset for this study are provided in Supplementary Table~\ref{sitab:2}. \ktwo and \tess photometry are publicly available in the Mikulski Archive for Space Telescopes (MAST; \url{https://archive.stsci.edu}). Images from the \spitzer Space Telescope are available in the \spitzer Heritage Archive (SHA; \url{https://sha.ipac.caltech.edu}), and images from the Las Cumbres Observatory (LCO) are available in the LCO Science Archive (\url{https://archive.lco.global}). All archival data can be found by searching for the stellar identifier V1298 Tau. Photometry from \spitzer, LCO, and other ground-based observations is available from the corresponding author upon reasonable request. 

\bmhead{Code Availability}
The transit light curve analyses were conducted using the \texttt{exoplanet} package, available on GitHub (\url{https://github.com/exoplanet-dev/exoplanet}). The stability analysis used the \texttt{SPOCK} package, which is also publicly available (\url{https://github.com/dtamayo/spock}). The $N$-body TTV analysis was performed using the publicly available \texttt{jnkepler} package (\url{https://github.com/kemasuda/jnkepler}), which includes an example notebook demonstrating the analysis for this system.

\bmhead{Acknowledgments} 
We thank the anonymous reviewers for their constructive comments.
This work is supported by JSPS KAKENHI Grant Numbers JP24H00017, JP24H00248, JP24K00689, JP24K17082, JP24K17083, JP25KJ0091, and JP25K17450, as well as by JSPS Grant-in-Aid for JSPS Fellows (Grant Number JP24KJ0241) and the JSPS Bilateral Program (JPJSBP120249910).
This work is based in part on observations made with the Spitzer Space Telescope, which was operated by the Jet Propulsion Laboratory, California Institute of Technology under a contract with NASA. Support for this work was provided by NASA through an award issued by JPL/Caltech.
This paper includes data collected by the Kepler mission and obtained from the MAST data archive at the Space Telescope Science Institute (STScI). Funding for the Kepler mission is provided by the NASA Science Mission Directorate. STScI is operated by the Association of Universities for Research in Astronomy, Inc., under NASA contract NAS 5–26555. 
This paper includes data collected with the TESS mission, obtained from the MAST data archive at the Space Telescope Science Institute (STScI). Funding for the TESS mission is provided by the NASA Explorer Program. STScI is operated by the Association of Universities for Research in Astronomy, Inc., under NASA contract NAS 5–26555.
These results are based on observations obtained with the Apache Point Observatory 3.5-meter telescope which is owned and operated by the Astrophysical Research Consortium. We wish to thank the APO 3.5m telescope operators in their assistance in obtaining these data.
Based in part on observations at the Kitt Peak National Observatory, NSF’s NOIRLab, managed by the Association of Universities for Research in Astronomy (AURA) under a cooperative agreement with the National Science Foundation. The WIYN 0.9m telescope is operated by WIYN Inc. on behalf of a Consortium of ten partner Universities and Organizations. WIYN is a joint facility of the University of Wisconsin–Madison, Indiana University, NSF's NOIRLab, the Pennsylvania State University, Purdue University, University of California, Irvine, and the University of Missouri. The authors are honored to be permitted to conduct astronomical research on Iolkam Du’ag (Kitt Peak), a mountain with particular significance to the Tohono O’odham.
Some of the data presented herein were obtained at the W. M. Keck Observatory, which is operated as a scientific partnership among the California Institute of Technology, the University of California and the National Aeronautics and Space Administration. The Observatory was made possible by the generous financial support of the W. M. Keck Foundation. The authors wish to recognize and acknowledge the very significant cultural role and reverence that the summit of Maunakea has always had within the indigenous Hawaiian community.  We are most fortunate to have the opportunity to conduct observations from this mountain.
This work makes use of observations from the Las Cumbres Observatory global telescope network.
This paper is based on observations made with the MuSCAT3 instrument, developed by Astrobiology Center and under financial supports by JSPS KAKENHI (JP18H05439) and JST PRESTO (JPMJPR1775), at Faulkes Telescope North on Maui, HI, operated by the Las Cumbres Observatory.
The research was carried out at the Jet Propulsion Laboratory, California Institute of Technology, under a contract with the National Aeronautics and Space Administration (80NM0018D0004).
C.C. acknowledges support by NASA Headquarters through an appointment to the NASA Postdoctoral Program at the Goddard Space Flight Center, administered by ORAU through a contract with NASA.
E.A.P. acknowledges support from Heising-Simons Foundation grant 2022-3833.

\bmhead{Author contributions}
J.H.L. planned and executed the observations, modelled the light curves, and performed the dynamical simulations. 
E.A.P., J.H.L, and K.M. analysed the TTVs. 
J.H.L., T.J.D., and E.A.P. led the writing of the manuscript. 
J.O. performed theoretical atmospheric evolution calculations.
D.N., K.B., and A.A.T. conducted dynamical analyses.
J.d.L., M.M., K.I., N.W., J.O.M., F.M., H.P., J.K., F.L., N.A.G., P.P.M.G., and E.P. conducted observations with MuSCAT and MuSCAT2.
N.N., A.F., and M.T. contributed LCO GTO time with which J.H.L. conducted MuSCAT3 observations.
A.Y. contributed Araki/ADLER observations.
A.R.-H. contributed LCO observations.
A.B. contributed KeplerCam observations.
E.E.M., D.R.C., V.G., L.A.H., L.M.R., E.R.N., A.W.M., A.V., K.T., J.-M.D., and L.P. contributed \spitzer observations.
G.S. and S.M. contributed APO observations. 
G.S., S.M., C.C., J.N., and J.H. contributed HDI observations. J.dL. performed photometric extractions of the MuSCAT, MuSCAT2, and MuSCAT3 data.
G.S. performed photometric extractions of the APO, HDI, and TMMT data.
All authors reviewed and approved the final manuscript.

\bmhead{Author Information} The authors declare no competing interests. Supplementary Information is available for this paper. Correspondence and requests for materials should be addressed to John Livingston (\url{john.livingston@nao.ac.jp}), Erik Petigura (\url{petigura@astro.ucla.edu}), and Trevor David (\url{trevorjdavid@gmail.com}).


\clearpage
\setcounter{figure}{0}
\renewcommand{\figurename}{Extended Data Fig.}
\renewcommand{\theHfigure}{ED.\thefigure}

\begin{figure}[ht!]
\centering
\includegraphics[width=\linewidth]{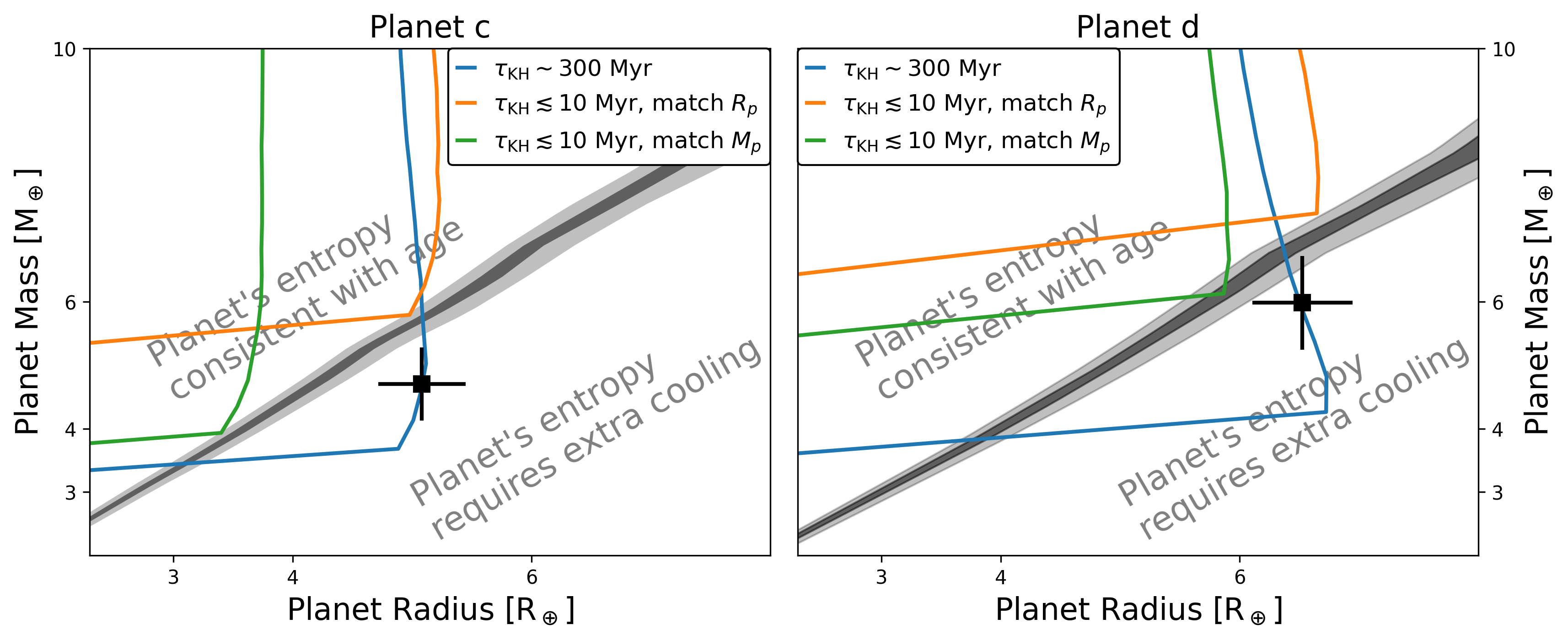}
\caption{
\textbf{Mass-radius diagram for \sname~c (left) and d (right) at the system's age of 23~Myr.} 
The gray shaded region marks the boundary of parameter space accessible by standard, high-entropy formation models (initial Kelvin-Helmholtz cooling timescale $\tau_\mathrm{KH}\lesssim$10~Myr), with the 1-$\sigma$ and 2-$\sigma$ uncertainty on its location shown in dark and light gray. Illustrative model tracks are shown for comparison: the green and orange lines demonstrate that high-entropy models fail to simultaneously match the measured mass and radius, while the blue line shows a successful low-entropy model ($\tau_\mathrm{KH} \sim 300$~Myr) consistent with the data. The positions of planets c and d firmly in the lower-right region provide direct evidence that their formation required a low-entropy initial state, such as that produced by a `boil-off' phase.
}
\label{edfig:1}
\end{figure}

\clearpage

\begin{figure}[ht!]
\centering
\includegraphics[width=\linewidth]{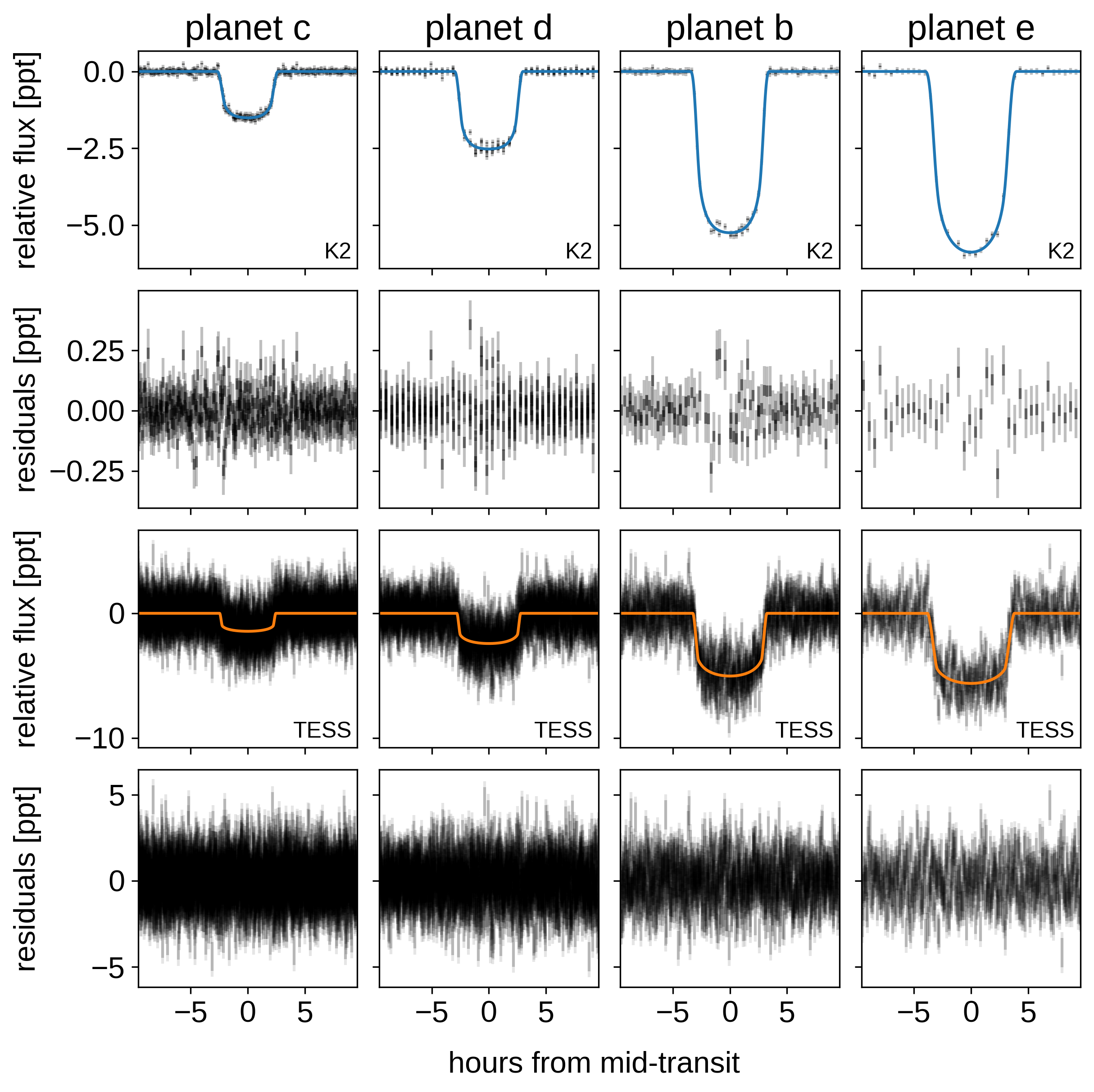}
\caption{\textbf{\ktwo and \tess data with models from the joint fit.} 
The top two rows show the variability-corrected K2 light curves centered on the transit of each planet with the best-fit transit model over-plotted, with the residuals from the fit shown below; dark error bars show the measured photometric uncertainties, while the lighter error bars denote the error bars including the jitter value from the fit. The bottom two rows show the same, but for the TESS data.
}
\label{edfig:2}
\end{figure}

\clearpage

\begin{figure}[ht!]
\centering
\includegraphics[width=\linewidth]{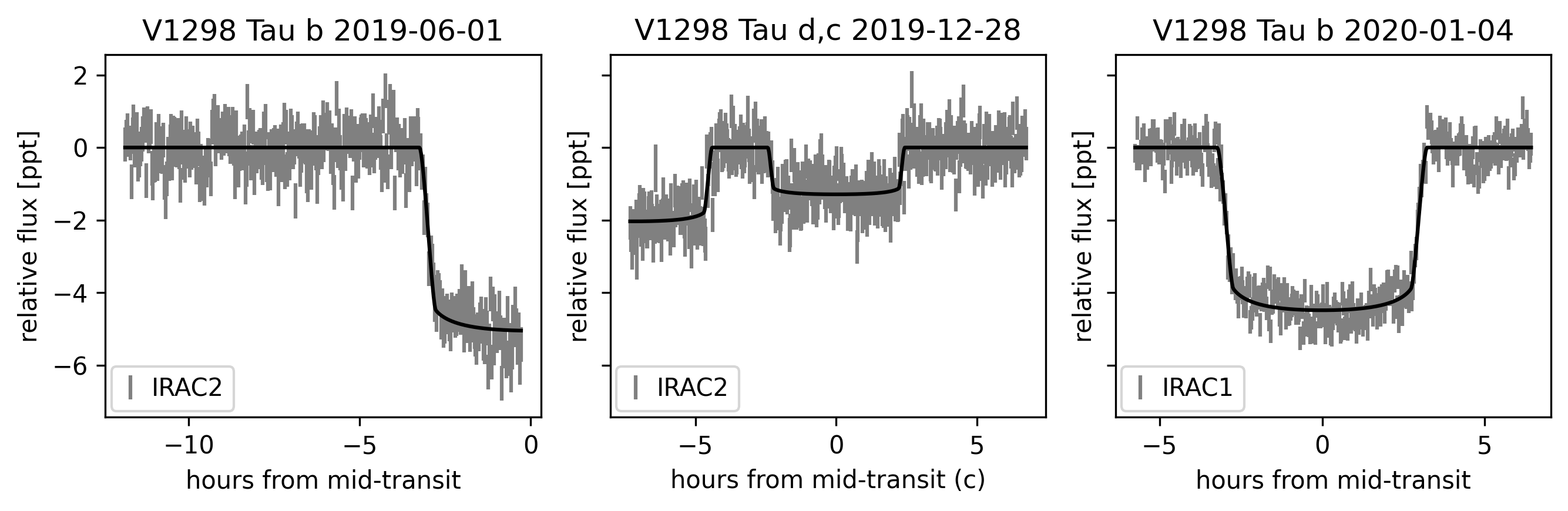}
\caption{\textbf{\spitzer observations used in this work with best-fit transit models.} The photometry has been corrected for stellar variability and systematics.}
\label{edfig:3}
\end{figure}

\clearpage

\begin{figure}[ht!]
\centering
\includegraphics[width=\linewidth]{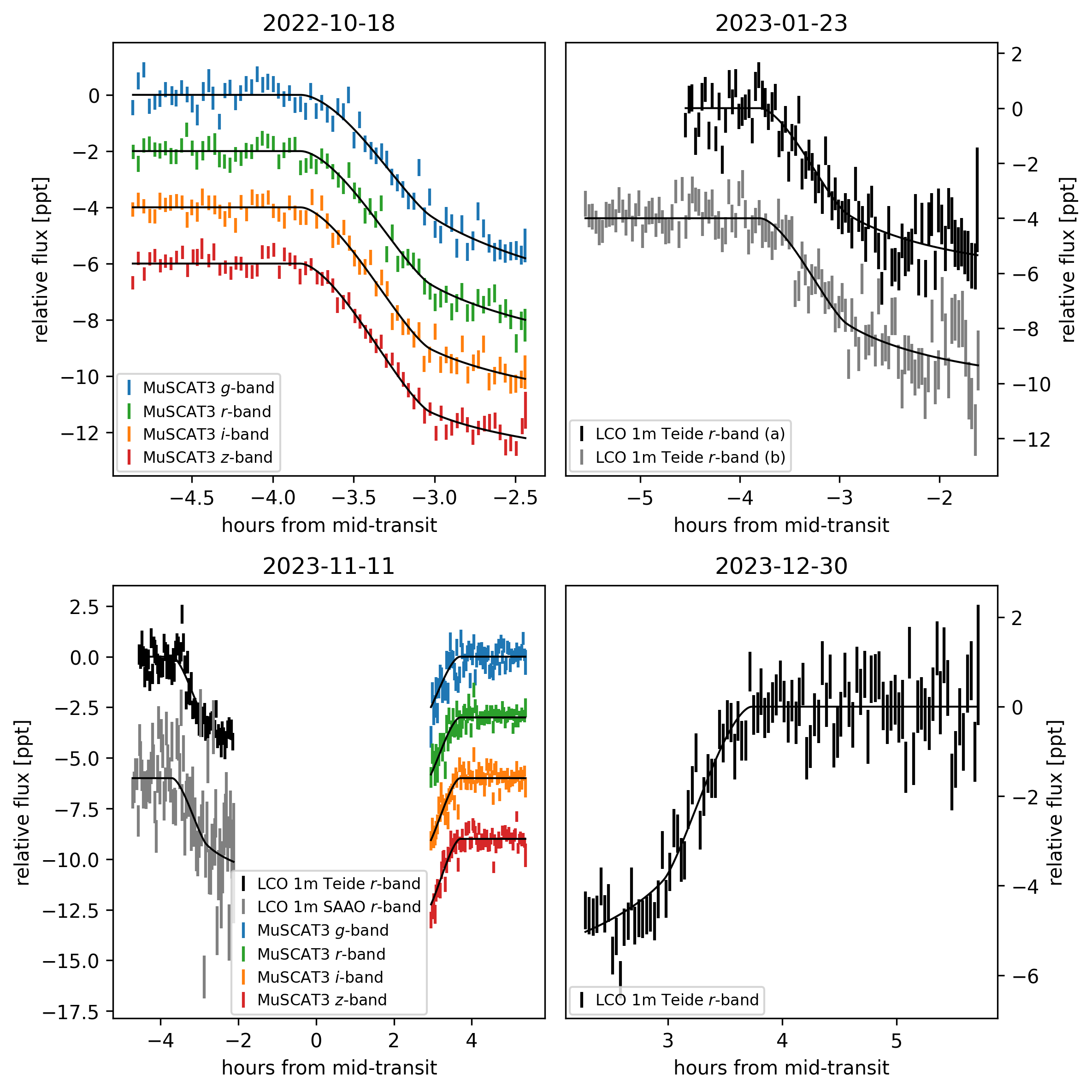}
\caption{\textbf{Recovery of \sname~e from ground-based transit observations.} 
The top left panel shows the partial transit recovered on 2022 October 18, which resolved the period ambiguity. The remaining panels show additional follow-up observations.}
\label{edfig:4}
\end{figure}

\clearpage

\begin{figure}[ht!]
\centering
\includegraphics[width=\linewidth]{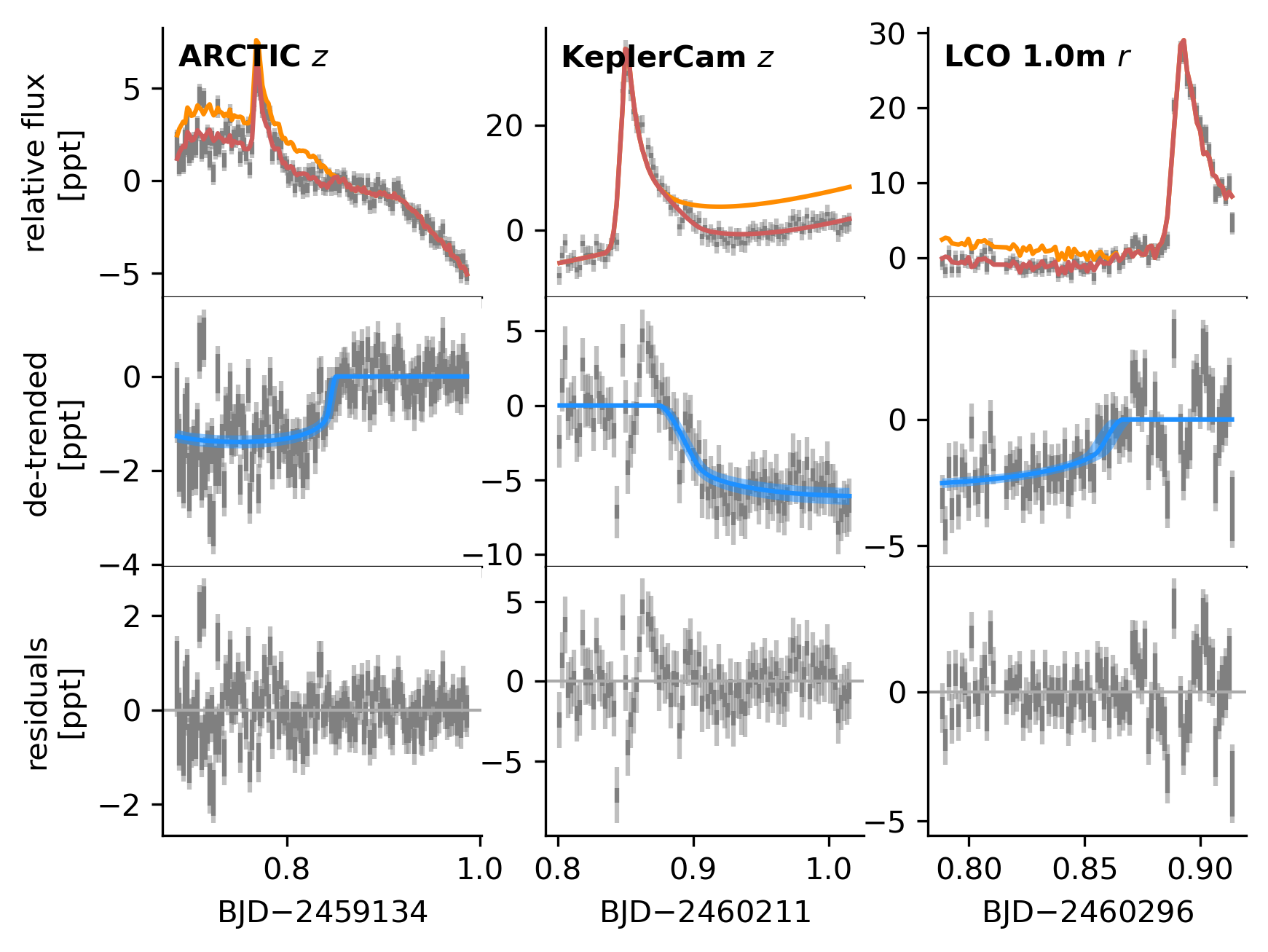}
\caption{\textbf{Datasets containing flares:} a transit of planet c from ARCTIC (left); a transit of planet e from KeplerCam (middle); a transit of planet d from LCO (right). The residuals from the fit to the KeplerCam data have a feature that may be due to a smaller, secondary flare.}
\label{edfig:5}
\end{figure}

\clearpage

\begin{figure}[ht!]
\centering
\includegraphics[width=\linewidth]{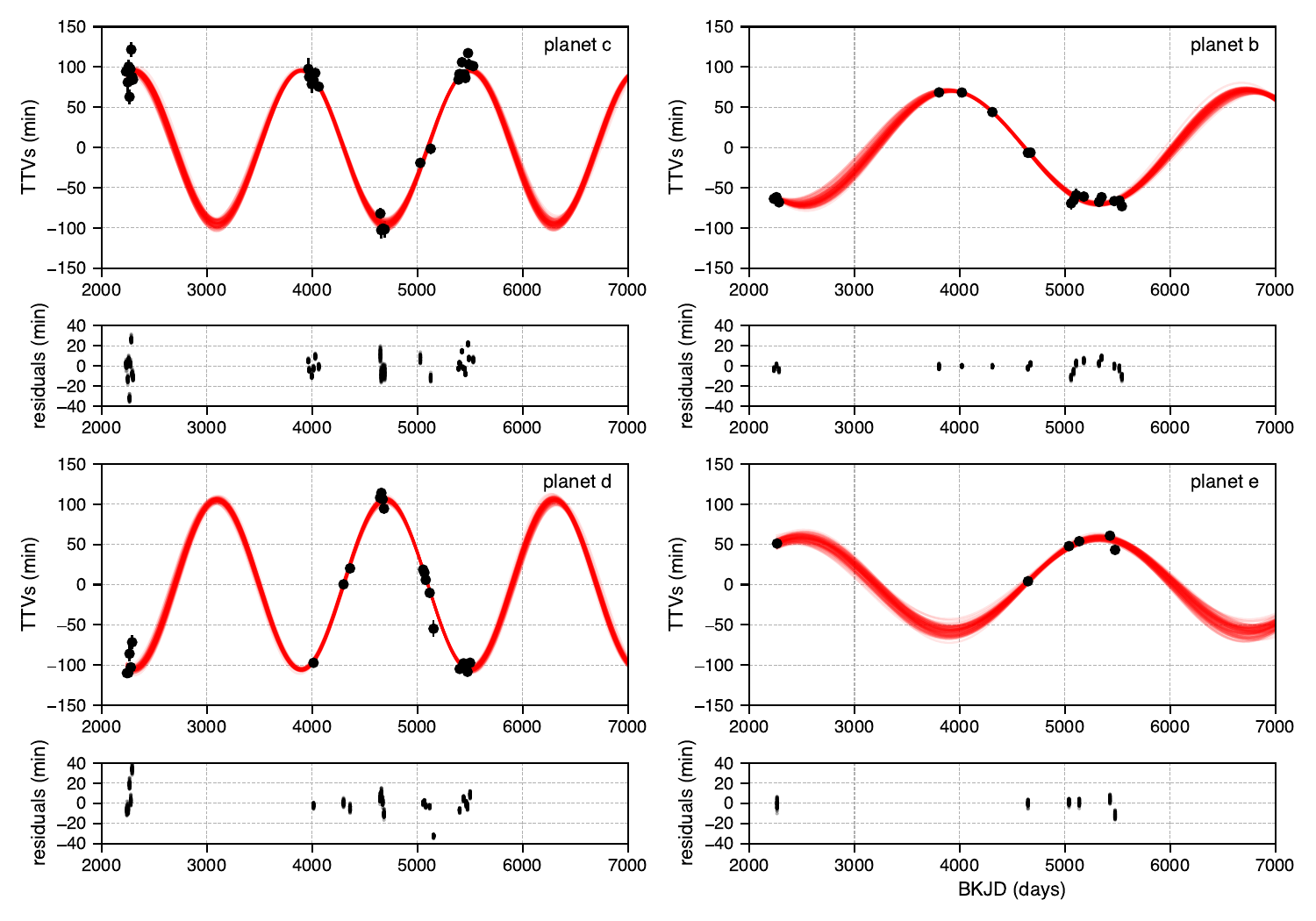}
\caption{\textbf{Multi-harmonic model of \sname TTVs described in Methods.} 
The top left panel shows the measured deviations of planet c's transit times from the best-fitting linear ephemeris. The red lines show draws from credible models. The second panel shows the difference between measured times and the model predictions. Other panels show the same quantities for planets d, b and e. The TTVs of planets c and d are dominated by interactions between the two planets. They are well-described by a sinusoid with an amplitude of $\sim$100~min and a period of $\isr{P}{cd} = 1604 \pm 12$~days. The TTVs of planets b and e are dominated by interactions between the two planets. Here, the TTVs are well-described by a sinusoid with an amplitude of $\sim$50~min and a period of $2852 \pm 50$~days.
}
\label{edfig:6}
\end{figure}

\clearpage

\begin{figure}[ht!]
\centering
\includegraphics[width=\linewidth]{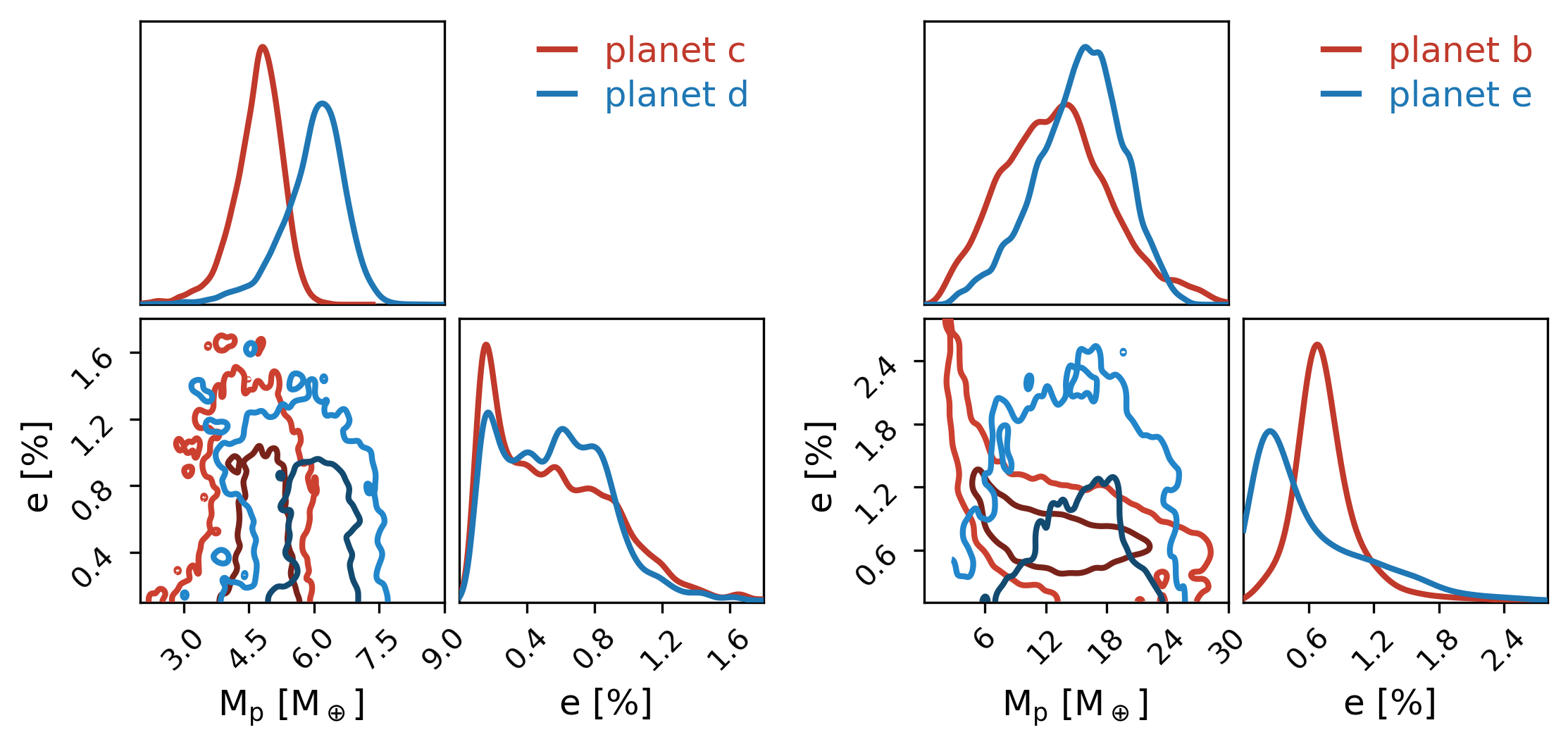}
\caption{\textbf{Mass and eccentricity posterior distributions.} Left: joint and marginal posterior distributions of mass and eccentricity for planets c and d. Right: same as left but for planets b and e. The joint posterior panels show 1- and 2-$\sigma$ contours for each planet.}
\label{edfig:7}
\end{figure}

\clearpage

\begin{figure}[ht!]
\centering
\includegraphics[width=\linewidth]{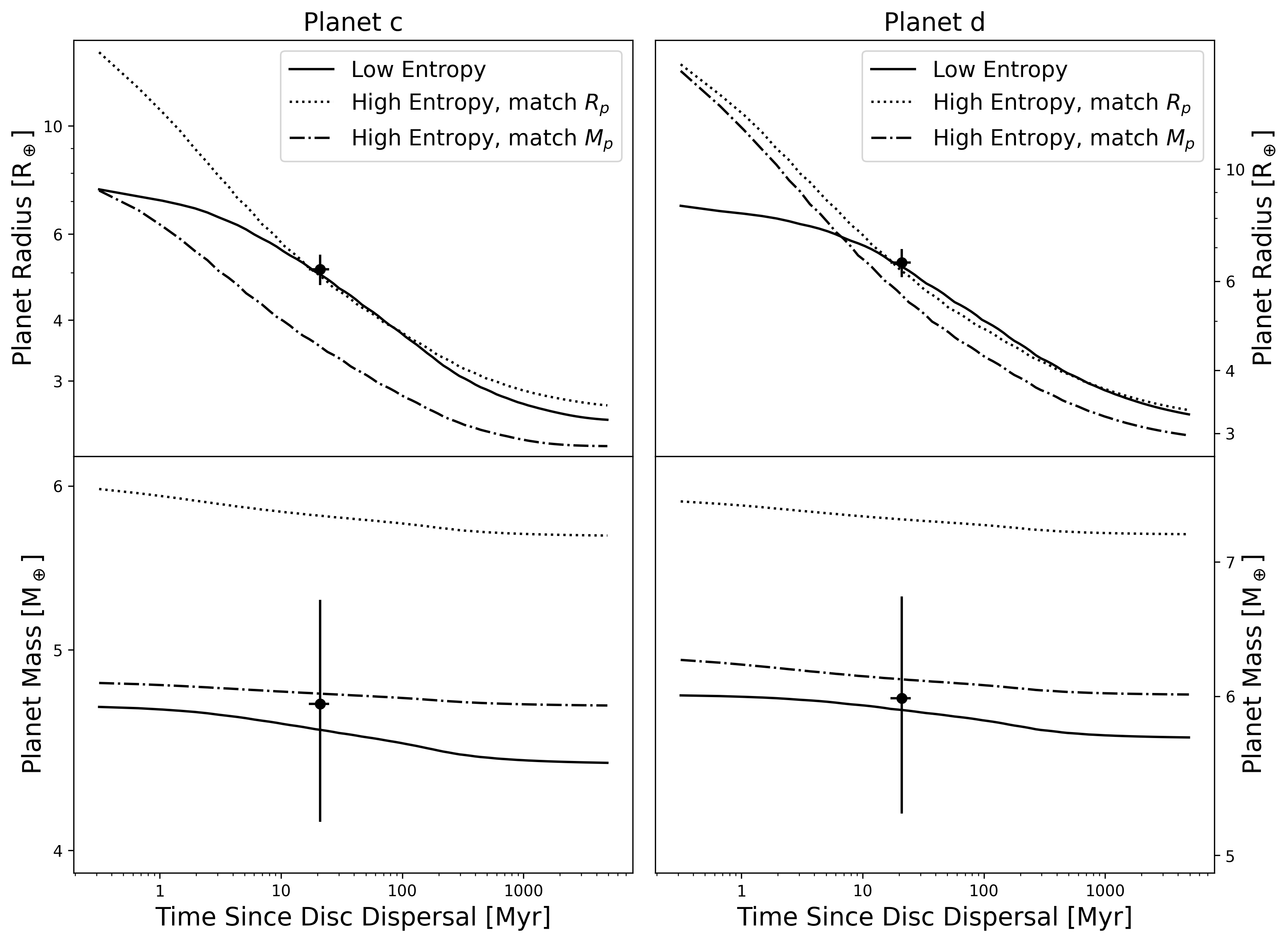}
\caption{\textbf{Planetary evolution models for \sname c and d.} 
Left: plausible and implausible mass/radius evolution tracks of planet c. We ran a grid of planetary models with different masses, initial envelope fractions, and envelope entropies. The black line best matches the present-day mass and radius together, but requires low entropy. If we require high entropy (orange curve) we may match the radius, cannot match the mass. Similarly, we may match mass (blue curve), but cannot match the radius. Therefore a self-consistent formation model requires a low initial entropy. The planet evolves into a sub-Neptune of $\approx2.5\Re$. Right: same as left, but for planet d, which matures into a slightly larger sub-Neptune.}
\label{edfig:8}
\end{figure}

\clearpage

\setcounter{table}{0}
\renewcommand{\tablename}{Extended Data Table}
\renewcommand{\theHtable}{ED.\thetable}

\begin{table}[ht]
\tiny
\centering
\caption{Results of \ktwo-\tess joint modeling.\label{edtab:1}}
\begin{tabular*}{\textwidth}{*{5}{@{\hskip 1.5pt}l@{\hskip 1.5pt}}}
\hline
{\it Star} & {\it Value} & & & \\
\hline
$u_1$ ($Kp$) &$0.11^{+0.13}_{-0.08}$ & & & \\
$u_2$ ($Kp$) &$0.71^{+0.13}_{-0.18}$ & & & \\
$u_1$ ($T$) &$0.24^{+0.21}_{-0.16}$ & & & \\
$u_2$ ($T$) &$0.24^{+0.23}_{-0.27}$ & & & \\
$\sigma_\mathrm{jit}$ ($Kp$) [ppt] &$0.10 \pm 0.01$ & & & \\
$\sigma_\mathrm{jit}$ ($T$) [ppt] &$0.85 \pm 0.01$ & & & \\
$\sigma_\mathrm{GP}$ ($Kp$) [ppt] &$16.0^{+1.7}_{-1.4}$ & & & \\
$\sigma_\mathrm{GP}$ ($T$) [ppt] &$13.0^{+1.5}_{-1.3}$ & & & \\
$\rho_\mathrm{GP}$ [days] &$3.86 \pm 0.30$ & & & \\
\hline
{\it Planet} & {\it c} & {\it d} & {\it b} & {\it e} \\
\hline
$R_p/R_\star$ & $0.0356^{+0.0018}_{-0.0015}$ & $0.0458 \pm 0.0015$ & $0.0661^{+0.0019}_{-0.0017}$ & $0.0716 \pm 0.0036$ \\
$b$ & $0.36 \pm 0.24$ & $0.21^{+0.21}_{-0.14}$ & $0.26 \pm 0.17$ & $0.61^{+0.06}_{-0.10}$ \\
$P$ [days] & $8.248714 \pm 0.000007$ & $12.402153 \pm 0.000008$ & $24.140420 \pm 0.000012$ & $48.677058 \pm 0.000041$ \\
$T_0$ [BKJD] & $3460.341238 \pm 0.000970$ & $3454.802142 \pm 0.000778$ & $3465.209601 \pm 0.000585$ & $3480.547375 \pm 0.001000$ \\
$T_{14}$ [hours] & $4.89^{+0.08}_{-0.06}$ & $5.57^{+0.06}_{-0.05}$ & $6.45^{+0.06}_{-0.05}$ & $7.56^{+0.12}_{-0.13}$ \\
\hline
\end{tabular*}
\end{table}

\begin{table}[ht]
\tiny
\begin{center}
\begin{minipage}{\textwidth}
\caption{Results of $N$-body TTV modeling}\label{edtab:2}
\begin{tabular*}{\textwidth}{*{5}{@{\hskip 5pt}l@{\hskip 5pt}}}
\toprule
Parameter & Mean & Standard Deviation & 95\% HDI & Prior \\
\midrule
\sname~c\\
$M_p/M_\star$ & $0.0000128$ & $0.0000015$ & $[0.0000100,0.0000154]$
 & $\mathcal{LU}$($0.0000030,0.0003003$)\\
$T_c$ (BKJD)&$2231.2835$&$0.0024$&$[2231.2789,2231.2882]$&$\mathcal{U}$($2231.2238,2231.3238$)\\
$P$ (days)&$8.24981$&$0.00039$&$[8.2490,8.2504]$&$\mathcal{U}$($8.1666,8.3316$)\\
$e\cos\omega$&$0.00074$&$0.00349$&$[-0.006,0.007]$&$e\sim \mathcal{LU}$($10^{-3},0.2$)\\
$e\sin\omega$&$0.0038$&$0.0041$&$[-0.003,0.012]$&$\omega\sim \mathcal{U}$($0,2\pi$)\\
\\
\sname~d\\
$M_p/M_\star$ & $0.0000163$ & $0.0000019$ & $[0.0000125,0.0000196]$ & $\mathcal{LU}$($0.0000030,0.0003003$)\\
$T_c$ (BKJD)&$2239.3916$&$0.0033$&$[2239.3850,2239.3979]$&$\mathcal{U}$($2239.3685,2239.4685$)\\
$P$ (days)&$12.40103$&$0.00066$&$[12.3999,12.4023]$&$\mathcal{U}$($12.2775,12.5256$)\\
$e\cos\omega$&$-0.0030$&$0.0029$&$[-0.009,0.002]$&$e\sim \mathcal{LU}$($10^{-3},0.2$)\\
$e\sin\omega$&$-0.0036$&$0.0035$&$[-0.010,0.003]$&$\omega\sim \mathcal{U}$($0,2\pi$)\\
\\
\sname~b\\
$M_p/M_\star$ & $0.0000358$ & $0.0000145$ & $[0.0000064,0.0000620]$ & $\mathcal{LU}$($0.0000030,0.0003003$)\\
$T_c$ (BKJD)&$2234.0507$&$0.0011$&$[2234.0486,2234.0534]$&$\mathcal{U}$($2234.0282,2234.1282$)\\
$P$ (days)&$24.13934$&$0.00030$&$[24.1387,24.1399]$&$\mathcal{U}$($23.8985,24.3813$)\\
$e\cos\omega$&$0.0048$&$0.0024$&$[-0.000,0.010]$&$e\sim \mathcal{LU}$($10^{-3},0.2$)\\
$e\sin\omega$&$0.0055$&$0.0045$&$[-0.001,0.014]$&$\omega\sim \mathcal{U}$($0,2\pi$)\\
\\
\sname~e\\
$M_p/M_\star$ & $0.0000418$ & $0.0000114$ & $[0.0000203,0.0000645]$ & $\mathcal{LU}$($0.0000030,0.0003003$)\\
$T_c$ (BKJD)&$2263.6241$&$0.0050$&$[2263.6145,2263.6340]$&$\mathcal{U}$($2263.5672,2263.6672$)\\
$P$ (days)&$48.68028$&$0.00037$&$[48.6795,48.6810]$&$\mathcal{U}$($48.1909,49.1645$)\\
$e\cos\omega$&$-0.00065$&$0.00677$&$[-0.016,0.013]$&$e\sim \mathcal{LU}$($10^{-3},0.2$)\\
$e\sin\omega$&$-0.00033$&$0.00594$&$[-0.012,0.012]$&$\omega\sim \mathcal{U}$($0,2\pi$)\\
\\
noise model\\
$\ln V$&$0.90$&$0.31$&$[0.3,1.5]$&$\mathcal{U}(-2, 10)$\\
$\ln\nu$&$2.5$&$1.0$&$[1.0,4.6]$&$\mathcal{U}(\ln(0.5), \ln(100)$\\
\botrule
\end{tabular*}
\footnotetext{BKJD refers to BJD$-$2454833. Orbital elements are defined at $\mathrm{BKJD}=2230$. $\mathcal{U}$ refers to a uniform distribution. $\mathcal{LU}$ refers to a log-uniform distribution.}
\end{minipage}
\end{center}
\end{table}


\newpage
\appendix

\section*{Supplementary Information}

This document contains supplementary notes and tables. The data for Supplementary Table 2 are provided inline for convenience and are also available as a separate machine-readable file from the publisher's website at \url{https://doi.org/10.1038/s41586-025-09840-z}. Supplementary Tables 1 and 3 are included inline.

\subsection*{Ground-based observations}

\subsubsection*{LCO}

The majority of our follow-up transit observations were obtained from 2020 to 2024 using the Las Cumbres Observatory global telescope network. We primarily used the Sinistro and MuSCAT3 instruments, mounted on the 1m and 2m telescopes, respectively, with occasional use of the SBIGSTL6303 camera on 0.4m telescopes. Exposure times ranged from 8-20 seconds across various filters (SDSS $g\prime$, $r\prime$, $i\prime$, and $z\prime_s$), with moderate defocus to improve the duty cycle.
Data were calibrated using the BANZAI pipeline and processed with a custom aperture photometry pipeline. Optimal light curves were selected by minimizing flux RMS, using combinations of typically five comparison stars and aperture radii between 10 to 30 pixels. Data points beyond the saturation limit and outliers with large centroid offsets were removed. 

\subsubsection*{ARCTIC}

On the UT nights of 2020 January 5, January 14, February 16, and October 12, we observed transits of \sname~b (first observation), and planet c (remaining observations), using the Astrophysical Research Consortium Telescope Imaging Camera on the 3.5m ARC telescope at Apache Point Observatory (APO). The ARCTIC data were obtained with the Engineered Diffuser available on the ARCTIC instrument capable of maintaining a stable point spread profile throughout the observations. All of the observations were obtained with the Semrock 867/30nm filter. The data were bias, dark and flat fielded and extracted using the \texttt{AstroImageJ} software. 
The October 12 observations were affected by a stellar flare during transit, which complicated the analysis, so we excluded it from our TTV analyses.

\subsubsection*{KeplerCam}

On UT 2019 November 17, we observed a full transit of \sname~c with the KeplerCam instrument mounted on the 1.2-meter telescope at the Fred Lawrence Whipple Observatory. KeplerCam is a single chip CCD with a Fairchild 486 detector. It has a field of view of 23.1' x 23.1' with a pixel scale of 0.672 arcsec/pixel in 2x2 binned mode. We observed the star using the SDSS $z^\prime$ filter with 8 second exposures. Aperture photometry was performed in \texttt{AstroImageJ} using a 10 pixel radius aperture and a sky annulus with inner and outer radii of 15 and 25 pixels, respectively.

\subsubsection*{HDI}

On the nights of UT 2019 November 16 and 2020 February 16, we obtained diffuser-assisted photometry of the transits of \sname~c using the Half Degree Imager, mounted on the 0.9-meter WIYN telescope at Kitt Peak. HDI has a $4096 \times 4096$ back-illuminated CCD from e2v covering a $29.2\arcmin \times 29.2\arcmin$ Field of View at a plate scale of $0.425\mathrm{\arcsec\:pixel^{-1}}$. To obtain high precision observations of the star, we used the Engineered Diffuser installed on the 0.9m WIYN Telescope. We observed the target in the SDSS $i^\prime$ filter in 1x1 binning with a gain of 1.3 e$^{-}$ / ADU, using an exposure time of $45\unit{s}$. To extract the photometry, we explored using a number of different apertures in \texttt{AstroImageJ}, where we ultimately adopted a reduction using a 18-20 pixel ($7.6-8.5\arcsec$) radius aperture and a sky annulus with inner and outer radii of 25 pixels ($11\arcsec$) and 40-50 pixels ($17-21\arcsec$), respectively, which resulted in the lowest overall photometric scatter. 

\subsubsection*{TMMT}

On the nights of UT 2019 November 8, November 16, and December 11, we observed transits of \sname~c using the Three-hundred MilliMeter Telescope at Las Campanas Observatory. The observations were performed slightly out of focus in the Cousins $I_C$ filter, resulting in a point spread function FWHM of $6-7\arcsec$. TMMT has a gain of 1.4 e$^{-}$ / ADU and a plate scale $1.2\mathrm{\arcsec\:pixel^{-1}}$ in the $1 \times 1$ binning mode used. For all observations, we used an exposure time of $30 \unit{s}$. To extract the photometry, we explored using a number of different apertures in \texttt{AstroImageJ}, where we ultimately adopted a reduction using a 7-8 pixel ($8.4-9.6\arcsec$) radius aperture and a sky annulus with inner and outer radii of 10 ($12\arcsec$) and 18-20 pixels ($22-24\arcsec$), respectively, which resulted in the lowest overall photometric scatter.

\subsubsection*{MuSCAT}

On the night of 2020 October 20 UT, we observed \sname with MuSCAT, mounted on the NAOJ 1.9m telescope located in Okayama, Japan. Data were calibrated using the standard instrument pipeline, including dark subtraction, flat fielding, and linearity correction. As was done for LCO data, light curves were produced using a custom aperture photometry pipeline.

\subsubsection*{MuSCAT2}

On the nights of 2022 Novemeber 3, 2023 October 5, 2023 November 7, 2023 December 2, 2023 December 10, and 2024 January 4, we observed \sname with MuSCAT2, mounted on the 1.5m TCS telescope (Telescopio Carlos Sanchez) at Teide Observatory, Tenerife. MuSCAT2 data calibration and photometry were performed in the same fashion as the MuSCAT data described above.

\subsubsection*{Koyama}

On 2020 December 9 UT, we observed a partial transit of \sname~d using the ADLER camera mounted on the 1.3m Araki Telescope telescope at Koyama Astronomical Observatory, Kyoto Sangyo University, Japan. ADLER has a Spectral Instruments 850 series 2048$\times$2048-pixel CCD with a pixel scale of 0.357 \arcsec/pix, yielding a $12 \arcmin \times 12 \arcmin$ FoV. A $\sim$4.7-hour sequence of 120 second exposures was obtained in $z\prime$-band from 08:39 to 13:18 UT, using moderate telescope defocus to avoid saturation of the target and comparison stars. On 2023 March 8 from 10:09 to 12:54 UT, and again on 2024 February 9 from 10:11 to 13:04 UT, we observed partial transits of \sname~b using the same instrumental setup.

\subsection*{Datasets containing flares}

We modeled datasets containing flares using our standard approach, augmented with a parametric flare model that characterizes flares by their peak time ($t_\mathrm{peak}$), amplitude ($a$), and full-width at half-maximum ($w$). Significant flares were observed in ARCTIC data (2020 October 12), KeplerCam data (2023 September 24), and LCO data (2023 December 18), with amplitudes ranging from 6 to 42 ppt and timescales of 14 to 21 minutes. These datasets and models are shown in Extended Data Fig.~\ref{edfig:5}.
The ARCTIC dataset yielded a flare with parameters $t_\mathrm{peak} = 2459134.7691 \pm 0.0003$ BJD, $a = 6.0 \pm 0.5$ ppt, and $w = 14.3 \pm 2.2$ minutes; the KeplerCam dataset yielded $t_\mathrm{peak} = 2460211.8506 \pm 0.0002$ BJD, $a = 42.0 \pm 1.2$ ppt, and $w = 21.1 \pm 1.4$ minutes; the LCO dataset yielded $2460296.8921 \pm 0.0002$ BJD, $a = 30.1 \pm 1.0$ ppt, and $w = 18.1 \pm 1.1$ minutes. These flare observations may prove valuable for future studies of the activity of \sname.

\subsection*{Mass constraints from analytic TTV modeling}

The four planets in the \sname system tug on one another due to gravity. These interactions result in transit-timing variations (TTVs) of several hours. Analytic models of TTVs have been developed in numerous previous works,
and several groups have developed $N$-body codes.
Many TTV studies have been performed for planets from the four-year \kepler mission. These studies had the benefit of near-continuous sampling over a full TTV period. The \sname dataset presents a different set of challenges due to low phase coverage and heterogeneous uncertainties. We thus analyzed the system using a mix of analytic and numerical models.

Lithwick et al. (2012; hereafter L12) developed an analytic model for TTVs due to pairs of planets near (but not too close to or in) resonance. Subsequently, Nesvorny et al. (2016; hereafter N16) developed a more general analytic model valid for planets both in and near resonance. The L12 model may be derived as a limiting case of the N16 model. We used both models, where appropriate. 

L12 showed that pairs of planets near resonance exhibit anti-correlated and sinusoidal TTVs. The TTV period (sometimes called the `super-period') is given by 
\begin{equation}
\isr{P}{12} = \frac{P_2}{ j \lvert \Delta \rvert}
\label{eqn:super-period}
\end{equation}
where $j$ defines the first-order resonance and $\Delta$ characterizes the proximity to it:
\begin{equation}
\Delta = \frac{P_2}{P_1}\frac{j - 1}{j} - 1.
\label{eqn:delta}
\end{equation}
We computed the proximity parameter and TTV period using a linear fit the measured transit times. We found 
$\Delta_\mathrm{cd}$ = $\val{delta-cd-per}$\%,
$\Delta_\mathrm{db}$ = $\val{delta-db-per}$\%, and
$\Delta_\mathrm{be}$ = $\val{delta-be-per}$\%.
The expected TTV periods are
\isr{P}{cd} = \val{ttv-super-per-cd-lin} days,
\isr{P}{db} = \val{ttv-super-per-db-lin} days, and
\isr{P}{be} = \val{ttv-super-per-be-lin} days.

The TTV amplitude is proportional to $1/\Delta$. Since, \isr{\Delta}{db} is much larger than both \isr{\Delta}{cd} and \isr{\Delta}{be} the TTVs of c and d will be dominated by c-d interactions and the TTVs of b and e will be dominated by b-e interactions. The b-d interactions will be second order corrections. Having thus separated the problem, we fit the following multi-harmonic model to the TTV timeseries. 
\begin{eqnarray}
\isr{t}{c} = \isr{t}{c,0} +  P_c \isr{i}{c} + \isr{A}{cd} \sin(2 \pi / \isr{P}{cd} (t - \isr{t}{cd,0}) ) \label{eqn:hm-c} \\
\isr{t}{d} = \isr{t}{d,0} +  P_d \isr{i}{d} + \isr{A}{dc} \sin(2 \pi / \isr{P}{cd} (t - \isr{t}{cd,0}) ) \label{eqn:hm-d} \\
\isr{t}{b} = \isr{t}{b,0} +  P_b \isr{i}{b} + \isr{A}{be} \sin(2 \pi / \isr{P}{be} (t - \isr{t}{be,0}) ) \label{eqn:hm-b} \\
\isr{t}{e} = \isr{t}{e,0} +  P_e \isr{i}{e} + \isr{A}{eb} \sin(2 \pi / \isr{P}{be} (t - \isr{t}{be,0}) ) \label{eqn:hm-e}
\end{eqnarray}
For each planet, a baseline linear ephemeris is specified by the first two terms, where $i$ is an integer that indexes each transit. Sinusoidal TTVs are specified by the last term. The parameters \isr{A}{cd}, \isr{P}{cd}, and \isr{t}{cd,0} specify the amplitude, period, and phase of the TTVs of planet c due to d. Equivalent parameters for other planets follow the same convention. Note that due to the anti-correlated nature of  TTVs, we set $\isr{P}{cd} = \isr{P}{db}$, $\isr{t}{cd,0} = \isr{t}{dc,0}$. An equivalent symmetry exists for the b-e pair. We modeled the transit times allowing the following 16 parameters to vary: \{\isr{t}{c,0}, \isr{P}{c}, \isr{A}{cd}, \isr{P}{cd}, \isr{t}{cd}, \isr{t}{d,0}, \isr{P}{d}, \isr{A}{dc}, \isr{t}{b,0}, \isr{P}{b}, \isr{A}{be}, \isr{P}{be}, \isr{t}{be}, \isr{t}{e,0}, \isr{P}{e}, \isr{A}{eb}\}. We explored credible models using MCMC. Extended Data Fig.~\ref{edfig:6} shows the measured transit times and draws from the set of credible models.

We found \isr{P}{be} to be $\val{ttv-per-be-mean} \pm \val{ttv-per-be-std}$~days. The b-e pair is well described by a sinusoid with a period matching the L12 prediction of $\val{ttv-super-per-be-lin}$~days, indicating that the L12 model is an adequate model of the TTV signal. In contrast, $\isr{P}{cd}$ is $\val{ttv-per-cd-mean} \pm \val{ttv-per-cd-std}$~days, which is inconsistent with the L12 predicted value of \val{ttv-super-per-cd-lin}~days at high significance. The L12 model is not appropriate for the c-d pair and the more general N16 model is needed.

\subsubsection*{TTV analysis of planets c and d}

Observed TTVs depend on variations in the position of the planet in its orbit and the orbit's orientation according to: 
\begin{equation}
\delta t = \frac{1}{n} (-\delta \lambda + 2 \delta h)
\end{equation}
where $n$ is the mean motion, $\lambda$ is the mean longitude, $h = e \sin\varpi$, and $\varpi$ is the longitude of periastron. When a system is sufficiently far from resonance, sinusoidal variations due to $\lambda$ and $h$ have the same period, $\isr{P}{ttv,\lambda} \approx \isr{P}{ttv,h}$. Sufficiently close to resonance, the two frequencies diverge and $P_\lambda < P_h$.

N16 showed that $\lambda$ oscillates with a period  

\begin{equation}
\isr{P}{ttv,\lambda} = \frac{P_1 + P_2}{2}
      \frac{P_\tau}{2 \pi}
      \left(\frac{m_1 + m_2}{M_\star}\right)^{-2/3}
      \left[\frac{3}{2}k(k-1)\lvert f_1 \rvert f_2\right]^{-1/3}
\end{equation}
where $k$ = 3 is the degree of resonance, $f_1 = -2.025$ and $f_2 = 2.484$ are order unity resonant coefficients of the disturbing function, and $P_\tau$ is the dimensionless period of resonant libations (typically $\approx  2-4$). Thus, $P_\lambda$ is sensitive to the sum of the planet masses. In addition, the TTV amplitudes are 
\begin{eqnarray}
A_1 & = & \frac{3(k - 1)}{\Lambda_1 \nu}\frac{P_\tau}{2 \pi} A_\Psi \\
A_2 & = & \frac{3k}{\Lambda_2 \nu}\frac{P_\tau}{2 \pi} A_\Psi
\end{eqnarray}
where $\nu = (3/2)[(k-1)^2 n_1 / \Lambda_1 + k^2 n_2 / \Lambda_2]$, $\Lambda_j = m_j \sqrt{G M_\star a_j}$, and $A_\Psi$ is the dimensionless amplitude of resonant librations. We do not know $A_\Psi$ in advance, which could range from 0 for an exact resonance to >1 for large librations. However, we note that $A_1 / A_2 \simeq -[(k-1)/k]^{2/3}M_2/M_1$, and thus the ratio of TTV amplitudes is closely related to the planet mass ratio. 

The TTV period and amplitudes together constrain both the mass ratio and the sum of the masses, which together are sufficient to constrain the masses of the individual planets. We did so through a simple, albeit brute-force, importance sampling scheme. We simulated $10^{10}$ planet pairs with properties drawn according to the following distributions:

\begin{eqnarray}
m_1 &\sim& \text{LogUniform}(10^{-1}, 10^2)~\Me \\
M_2/M_1 &\sim& \text{LogUniform}(0.5, 2) \\
e_1 \sim e_2 &\sim& \text{LogUniform}(10^{-4}, 10^{-1}) \\
\sigma_1 = k \lambda_2 - (k -1) \lambda_1 - \varpi_1 &\sim& U(0,2\pi) \\
\sigma_2 = k \lambda_2 - (k -1) \lambda_1 - \varpi_2 &\sim& U(0,2\pi) \\
\end{eqnarray}
We then applied the formulae in N16 to determine $\isr{A}{cd,sim}$, $\isr{A}{dc,sim}$, $\isr{P}{cd,sim}$ for each simulated planet pair. We then computed 

\begin{eqnarray}
    \chi^2 = \left(\frac{\isr{A}{cd,sim} - \isr{A}{cd}}{\sigma(\isr{A}{cd})}\right)^2
           + \left(\frac{\isr{A}{dc,sim} - \isr{A}{dc}}{\sigma(\isr{A}{dc})}\right)^2
           + \left(\frac{\isr{P}{cd,sim} - \isr{P}{cd}}{\sigma(\isr{P}{cd})}\right)^2
\end{eqnarray}
to quantify the degree to which each simulated system matched the observed system. Because the prior volume is large, and $\{\isr{A}{cd},\isr{A}{dc},\isr{P}{cd}\}$ have small fractional uncertainties, our sampling efficiency is low. The fraction of planet pairs with $\chi^2 < 10$ was $\sim 10^{-4}$. The likelihood of any given model given the data is $\mathcal{L} \propto e^{-\chi^2 / 2}$. We therefore weighted each sample by $\mathcal{L}$ to derive credible intervals for each parameter. We found $m_c = \val{attv-m1}_{\val{attv-m1-err1}}^{+\val{attv-m1-err2}}~\Me$. Not surprisingly, the mass ratios were measured with high precision $m_d / m_c = \val{attv-r2} \pm \val{attv-r2-err2}$. Thus, $m_d = \val{attv-m2}_{\val{attv-m2-err1}}^{+\val{attv-m2-err2}}~\Me$. Finally, we determined that eccentricities were no more than a few percent: $e_c < \val{attv-e1-p95}$ and $e_d < \val{attv-e2-p95}$ at 95\% confidence. 

Let us pause to reflect on the remarkable insights achieved from the measurement of just three quantities. We have determined that the inner two planets are only a few Earth masses, despite being $\approx6$ Earth-radii and have eccentricities of only a few percent. These analytic results serve as a sanity check on our adopted $N$-body model.

\subsubsection*{TTV analysis of planets b and e}
As stated previously, the b-e pair may be treated with the L12 formalism, which predicts anti-correlated, sinusoidal TTVs with a period and amplitudes given by:

\begin{eqnarray}
\isr{P}{ttv} & = &  \frac{P_{2}}{j \lvert\Delta\rvert}, \\
A_1 & = & P_1 \frac{\mu_2}{\pi k^{2/3} (j-1)^{1/3}\Delta} \left(-f - \frac{3}{2} \frac{\Zfreestar}{\Delta}\right) \label{eqn:V1}\\
A_2 & = & P_2 \frac{\mu_1}{\pi k \Delta}\left(-g + \frac{3}{2} \frac{\Zfreestar}{\Delta} \right) \label{eqn:V2},
\end{eqnarray}
respectively, where $\mu$ is the planet-star mass ratio, and $f$ and $g$ are order unity scalar coefficients that depend on $j$ and $\Delta$ and are given in L12. For the b-e pair, $f = -1.190$ and $g = 0.4284$, neglecting terms of $\mathcal{O}(\Delta)$. \Zfreestar is the complex conjugate of the following linear combination of the planets complex eccentricities:
\begin{equation}
\label{eqn:zfree}
\Zfree = f \zfree{1} + g\zfree{2},
\end{equation}
where 
\begin{equation}
z = e \cos \varpi + i e \sin \varpi.
\end{equation}
We note a few differences between the L12 and N16 TTV models. In the L12 limit, the TTV period does not depend on the planet masses, only the orbital periods; in contrast, in the N16 model it depends on the sum of the masses. The individual TTV amplitudes depend on the product of planet mass times a linear combination of the eccentricities. Thus, we anticipate good constraints on $M_2/M_1$ and the mass-eccentricity product, but not on individual masses or eccentricities. This `mass-eccentricity degeneracy' has been discussed extensively in the literature, e.g. Deck et al. (2015) and Hadden et al. (2016). Also in the L12 formalism, the TTV amplitude is treated as a complex value, where the magnitude encodes the amplitude of the signal and the phase encodes the shift of the TTV signal relative to the time when the line-of-conjunctions points at the observer. 

The amplitudes $\lvert A_1 \rvert$ and $\lvert A_2 \rvert$ are given as $A_{be}$ and $A_{eb}$ in Supplementary Table~\ref{sitab:3}. We observe a small phase shift $\phi = \val{l12ttv-phase} \pm \val{l12ttv-phase-err}$ radians which indicates the eccentricities are non-zero $\lvert \Zfree \vert > 0$. However, since the phase shift is small compared to unity, it is unlikely that $\lvert \Zfree \rvert \gg \Delta$. Therefore, $0 < \lvert \Zfree \vert \lesssim \Delta = 0.82$; thus, the individual eccentricities are a few percent or less. 

As in the previous section, we used importance sampling to measure the planet masses. We simulated $10^8$ planet pairs with properties drawn according to the following distributions:

\begin{eqnarray}
\mu_1    &\sim & \text{LogUniform}(10^{0}, 10^2)~\Me / \Msun \\
M_2/M_1  &\sim & \text{LogUniform}(0.2, 5) \\
\lvert \Zfree \rvert   &\sim & \text{LogUniform}(10^{-4}, 10^{-1}) \\
\text{phase}(\Zfree)         &\sim & \text{Uniform}(0,2 \pi)
\end{eqnarray}
We then applied the formulae in L12 to determine $\isr{A}{be,sim}$, $\isr{A}{eb,sim}$, and $\isr{\phi}{be}$ for each simulated planet pair. We then computed 

\begin{eqnarray}
    \chi^2 = \left(\frac{\isr{A}{be,sim} - \isr{A}{be}}{\sigma(\isr{A}{be})}\right)^2
           + \left(\frac{\isr{A}{eb,sim} - \isr{A}{eb}}{\sigma(\isr{A}{eb})}\right)^2
           + \left(\frac{\isr{\phi}{be,sim} - \isr{\phi}{be}}{\sigma(\isr{\phi}{be})}\right)^2
\end{eqnarray}
The samples were weighted following the procedure in the previous section. We found $M_b = \val{l12ttv-m1e}_{\val{l12ttv-m1e-err1}}^{+\val{l12ttv-m1e-err2}}$~\Me, $M_e = \val{l12ttv-m2e}_{\val{l12ttv-m2e-err1}}^{+\val{l12ttv-m2e-err2}}$~\Me, and $\log\lvert \Zfree \rvert = \val{l12ttv-logZfreemag}_{\val{l12ttv-logZfreemag-err1}}^{+\val{l12ttv-logZfreemag-err2}}$. The large fractional mass uncertainties stem from the mass-eccentricity degeneracy discussed earlier. One may decrease the mass of planets b and e (keeping their ratio fixed) and increase \Zfree and produce the same TTV curve. Increased phase coverage of planet e's TTV curve will not dramatically improve the mass measurements. However, there are a number of possible avenues to improve these measurements. As shown in Fig. 1 of the main article, L12 interactions between planet d and b lead to a low-amplitude $\sim$500~day TTV period. Better constraining this signal will improve $M_d/M_b$ and hence $M_b$ and $M_e$. Secondary eclipse times directly measure $e\cos \omega$ (Winn 2010) and break the mass-eccentricity degeneracy. Exceptionally precise timing measurements of consecutive transits could resolve synodic chopping which does not suffer from the mass-eccentricity degeneracy (Deck et al., 2015).

\subsection*{Mass constraints from $N$-body TTV modeling}

Our $N$-body model computes transit times of each planet considering gravitational interactions between the planets and the star. The interaction was assumed to be Newtonian, and the light travel time was ignored. The model parameters are four planets' masses relative to that of the host star, and the osculating orbital elements (period $P$, eccentricity $e$, argument of periastron $\omega$, and time of inferior conjunction $T_c$) defined at the start of integration, $\mathrm{BKJD}=2230$, where $T_c$ is converted to the time of periastron passage $\tau$ via $2\pi (T_c-\tau)/P = E_0 - e\sin E_0$ with $E_0=2\arctan\left[\sqrt{{1-e}\over{1+e}}\tan\left({\pi \over 4}-{\omega \over 2}\right)\right]$. We assume coplanar orbits, and fix the orbital inclinations to be $\pi/2$ and longitudes of the ascending node to be $0$ for all the planets. In defining the orbital elements, we choose the sky plane to be the reference plane, and adopt the $+Z$-axis pointing toward the observer. 

The masses and osculating orbital elements are converted to Jacobi coordinates, using the total interior mass as the mass entering in the conversion from periods to coordinates.
An $N$-body integration is performed using a symplectic integrator with a fixed time step of 0.3~days (i.e., $\approx 1/27$ of the innermost period), which results in a typical fractional energy error of $\mathcal{O}(10^{-8})$. The resulting orbits are then used to derive transit times of the planets following a standard iterative scheme, in which a fourth-order Hermite integrator is used. The $N$-body code is implemented in {\tt JAX} to enable automatic differentiation with respect to the input orbital elements and mass ratios, and is available through GitHub as a part of the {\tt jnkepler} package.

When sampling from the posterior probability distribution, we adopt the log-likelihood function given by the Student's $t$ distribution:
\begin{equation}
    \ln \mathcal{L}(\theta)
    = \sum_i\left\{-{\nu+1 \over 2}\ln\left[1 + {(t_i- m_i(\theta))^2\over \nu V\sigma_i^2}\right]
    -{1\over 2}\ln(\pi\nu V \sigma_i^2) + \ln\Gamma\left({\nu+1}\over 2\right) - \ln\Gamma\left(\nu\over2\right)\right\},
\end{equation}
where $m_i(\theta)$ is the model, $\nu$ is the number of degrees of freedom, $V$ is the scale of the distribution, and $\Gamma(x)$ is the Gamma function.
The prior is assumed to be separable for each parameter and is listed in Extended Data Table~\ref{edtab:2}. In practice, the sampling was performed assuming a uniform prior distribution for $e$, i.e. the results were calculated from posterior samples that were resampled with weights proportional to $1/e$.
The sampling was performed using the No-U-Turn Sampler as implemented in {\tt NumPyro}. 
We ran four chains in parallel for 2,000 steps. The resulting chains had a split Gelman-Rubin statistic of $\hat{R}<1.02$, and the estimated number of effective samples was at least 500 for each parameter.
The mass and eccentricity posterior distributions are shown in Extended Data Fig.~\ref{edfig:7}.



\setcounter{table}{0}
\renewcommand{\tablename}{Supplementary Table}
\renewcommand{\theHtable}{SI.\thetable}

\begin{table}[ht]
\tiny
\begin{center}
\begin{minipage}{\textwidth}
\caption{Summary of transit observations analyzed in this work.}\label{sitab:1}
\begin{tabular*}{\textwidth}{*{6}{@{\hskip 1.5mm}l@{\hskip 1.5mm}}}
\toprule
Date             & Date             & Planet    & Instrument          & Filter                                    & Exposure    \\
Start            & End              &           &                     &                                           & Time        \\
(UT)             & (UT)             &           &                     &                                           & (s)         \\
\midrule   
2015-02-08 07:13 & 2015-04-20 04:17 & b,c,d,e   & \kepler/\ktwo       & \kepler bandpass                          & 1800        \\
2019-06-01 05:59 & 2019-06-01 17:31 & b         & \spitzer            & IRAC2                                     & 2           \\
2019-11-09 02:56 & 2019-11-09 06:31 & c         & TMMT                & Cousins I                                 & 30          \\
2019-11-17 04:13 & 2019-11-17 08:05 & c         & TMMT                & Cousins I                                 & 30          \\
2019-11-17 04:39 & 2019-11-17 10:28 & c         & HDI/WIYN0.9m        & SDSS $i^\prime$                           & 45          \\
2019-11-17 05:25 & 2019-11-17 12:08 & c         & KeplerCam           & SDSS $z^\prime$                           & 8           \\
2019-12-12 00:42 & 2019-12-12 00:42 & c         & TMMT                & Cousins I                                 & 30          \\
2019-12-28 07:03 & 2019-12-28 21:03 & c,d       & \spitzer            & IRAC2                                     & 2           \\
2020-01-04 17:42 & 2020-01-05 06:24 & b         & \spitzer            & IRAC1                                     & 0.4         \\
2020-01-05 01:45 & 2020-01-05 07:04 & b         & ARCTIC/ARC3.5m      & Semrock 857/30                            & 30          \\
2020-01-14 02:04 & 2020-01-14 07:07 & c         & ARCTIC/ARC3.5m      & Semrock 857/30                            & 30          \\
2020-02-16 01:44 & 2020-02-16 05:58 & c         & ARCTIC/ARC3.5m      & Semrock 857/30                            & 30          \\
2020-02-16 02:23 & 2020-02-16 06:22 & c         & HDI/WIYN0.9m        & SDSS $i^\prime$                           & 45          \\
2020-10-08 14:14 & 2020-10-08 18:10 & d         & LCO1m/Sinistro@SSO  & $z_s$                                     & 30          \\
2020-10-12 04:28 & 2020-10-12 11:40 & c         & ARCTIC/ARC3.5m      & Semrock 857/30                            & 30          \\
2020-10-20 12:47 & 2020-10-20 20:29 & b,c       & Okayama1.9m/MuSCAT              & $g\prime$,$r\prime$,$z\prime_s$           & 60,40,60   \\
2020-12-09 08:39 & 2020-12-09 13:18 & d         & Koyama1.3m/ADLER    & $z_s$                                     & 120         \\
2021-09-16 16:04 & 2021-11-05 22:41 & b,c,d,e   & \tess               & \tess bandpass                            & 120         \\
2022-10-07 23:38 & 2022-10-08 06:15 & c         & LCO1m/Sinistro@TO   & $r\prime$                                 & 15          \\
2022-10-18 11:55 & 2022-10-18 15:29 & e         & LCO40cm/SBIG@HO    & Johnson-Cousins $V$                       & 12          \\
2022-10-18 13:08 & 2022-10-18 15:35 & e         & LCO2m/MuSCAT3@HO    & $g\prime$,$r\prime$,$i\prime$,$z\prime_s$ & 8,8,12,20   \\
2022-11-03 23:03 & 2022-11-04 05:06 & d         & TCS/MuSCAT2@TO      & $g\prime$,$r\prime$,$i\prime$,$z\prime_s$ & 20,15,20,20 \\
2022-11-07 23:41 & 2022-11-08 03:55 & b         & LCO40cm/SBIG@TO     & $r\prime$                                 & 90          \\
2022-11-16 08:00 & 2022-11-16 09:25 & d         & LCO2m/MuSCAT3@HO    & $g\prime$,$r\prime$,$i\prime$,$z\prime_s$ & 8,8,12,20   \\
2022-11-28 21:59 & 2022-11-29 02:17 & d         & LCO1m/Sinistro@TO   & $r\prime$                                 & 15          \\
2022-11-28 23:03 & 2022-11-29 03:42 & d         & TCS/MuSCAT2@TO      & $g\prime$,$r\prime$,$i\prime$,$z\prime_s$ & 8,8,5,5     \\
2022-12-02 03:40 & 2022-12-02 07:14 & b         & LCO1m/Sinistro@CTIO & $r\prime$                                 & 15          \\
2022-12-02 03:54 & 2022-12-02 06:52 & b         & LCO1m/Sinistro@MO   & $r\prime$                                 & 15          \\
2022-12-02 04:12 & 2022-12-02 07:04 & b         & LCO1m/Sinistro@MO   & $r\prime$                                 & 15          \\
2022-12-26 00:59 & 2022-12-26 04:50 & b         & LCO1m/Sinistro@CTIO & $r\prime$                                 & 15          \\
2023-01-05 03:09 & 2023-01-05 08:05 & d         & LCO1m/Sinistro@MO   & $r\prime$                                 & 15          \\
2023-01-05 03:38 & 2023-01-05 07:34 & d         & LCO1m/Sinistro@MO   & $r\prime$                                 & 15          \\
2023-01-15 00:54 & 2023-01-15 03:04 & c         & LCO1m/Sinistro@CTIO & $r\prime$                                 & 15          \\
2023-01-15 00:54 & 2023-01-15 03:33 & c         & LCO1m/Sinistro@CTIO & $r\prime$                                 & 15          \\
2023-01-15 01:20 & 2023-01-15 08:03 & c         & LCO1m/Sinistro@MO   & $r\prime$                                 & 15          \\
2023-01-23 21:08 & 2023-01-24 01:02 & e         & LCO1m/Sinistro@TO   & $r\prime$                                 & 15          \\
2023-01-23 22:08 & 2023-01-24 01:01 & e         & LCO1m/Sinistro@TO   & $r\prime$                                 & 15          \\
2023-02-11 01:38 & 2023-02-11 04:25 & d         & LCO1m/Sinistro@MO   & $r\prime$                                 & 20          \\
2023-02-11 01:38 & 2023-02-11 06:00 & d         & LCO1m/Sinistro@MO   & $r\prime$                                 & 20          \\
2023-03-08 10:09 & 2023-03-08 12:54 & b         & Koyama1.3m/ADLER    & $z_s$                                     & 120         \\
2023-07-31 13:16 & 2023-07-31 15:02 & b         & LCO2m/MuSCAT3@HO    & $g\prime$,$r\prime$,$i\prime$,$z\prime_s$ & 8,8,12,20   \\
2023-08-24 17:05 & 2023-08-24 19:13 & b         & LCO1m/Sinistro@CTIO & $r\prime$                                 & 20          \\
2023-08-24 17:05 & 2023-08-24 19:13 & b         & LCO1m/Sinistro@CTIO & $r\prime$                                 & 20          \\
2023-10-06 01:07 & 2023-10-06 06:05 & c         & TCS/MuSCAT2@TO      & $g\prime$,$r\prime$,$i\prime$,$z\prime_s$ & 10,12,12,15 \\
2023-10-06 01:08 & 2023-10-06 04:03 & c         & LCO1m/Sinistro@TO   & $r\prime$                                 & 20          \\
2023-10-06 05:52 & 2023-10-06 07:36 & c         & LCO1m/Sinistro@CTIO & $r\prime$                                 & 20          \\
2023-10-06 05:57 & 2023-10-06 07:36 & c         & LCO1m/Sinistro@CTIO & $r\prime$                                 & 20          \\
2023-10-06 06:08 & 2023-10-06 07:36 & c         & LCO1m/Sinistro@CTIO & $r\prime$                                 & 20          \\
2023-10-14 06:53 & 2023-10-14 09:19 & c         & LCO1m/Sinistro@MO   & $r\prime$                                 & 20          \\
2023-10-14 07:38 & 2023-10-14 09:34 & c         & LCO1m/Sinistro@MO   & $r\prime$                                 & 20          \\
2023-10-17 01:53 & 2023-10-17 03:26 & d         & LCO1m/Sinistro@TO   & $r\prime$                                 & 20          \\
2023-10-17 06:53 & 2023-10-17 09:50 & d         & LCO1m/Sinistro@MO   & $r\prime$                                 & 20          \\
2023-10-17 07:08 & 2023-10-17 09:34 & d         & LCO1m/Sinistro@MO   & $r\prime$                                 & 20          \\
2023-10-22 13:09 & 2023-10-22 15:35 & c         & LCO2m/MuSCAT3@HO    & $g\prime$,$r\prime$,$i\prime$,$z\prime_s$ & 8,8,12,20   \\
2023-11-08 00:34 & 2023-11-08 05:52 & c         & TCS/MuSCAT2@TO      & $g\prime$,$r\prime$,$i\prime$,$z\prime_s$ & 5,5,5,5     \\
2023-11-11 23:40 & 2023-11-12 02:16 & e         & LCO1m/Sinistro@SAAO & $r\prime$                                 & 20          \\
2023-11-11 23:50 & 2023-11-12 02:15 & e         & LCO1m/Sinistro@TO   & $r\prime$                                 & 20          \\
2023-11-11 07:20 & 2023-11-12 09:46 & e         & LCO2m/MuSCAT3@HO    & $g\prime$,$r\prime$,$i\prime$,$z\prime_s$ & 8,8,12,20   \\
2023-11-16 07:42 & 2023-11-16 10:58 & c         & LCO2m/MuSCAT3@HO    & $g\prime$,$r\prime$,$i\prime$,$z\prime_s$ & 8,8,12,20   \\
2023-11-16 12:26 & 2023-11-16 14:00 & c         & LCO2m/MuSCAT3@HO    & $g\prime$,$r\prime$,$i\prime$,$z\prime_s$ & 8,8,12,20   \\
2023-11-23 12:40 & 2023-11-23 14:05 & d         & LCO2m/MuSCAT3@HO    & $g\prime$,$r\prime$,$i\prime$,$z\prime_s$ & 8,8,12,20   \\
2023-12-02 22:34 & 2023-12-03 03:36 & c         & TCS/MuSCAT2@TO      & $g\prime$,$r\prime$,$i\prime$,$z\prime_s$ & 11,12,11,11 \\
2023-12-11 01:43 & 2023-12-11 04:03 & c         & TCS/MuSCAT2@TO      & $g\prime$,$r\prime$,$i\prime$,$z\prime_s$ & 7,7,4,4     \\
2023-12-11 02:10 & 2023-12-11 10:30 & c         & KeplerCam           & SDSS $z^\prime$                           & 30          \\
2023-12-18 01:44 & 2023-12-18 07:26 & d         & KeplerCam           & SDSS $z^\prime$                           & 30          \\
2023-12-23 02:54 & 2023-12-23 05:50 & b         & LCO1m/Sinistro@MO   & $r\prime$                                 & 20          \\
2023-12-23 02:55 & 2023-12-23 06:50 & b         & LCO1m/Sinistro@MO   & $r\prime$                                 & 20          \\
2023-12-30 22:39 & 2023-12-31 02:05 & e         & LCO1m/Sinistro@TO   & $r\prime$                                 & 20          \\
2024-01-04 19:49 & 2024-01-04 22:52 & c         & TCS/MuSCAT2@TO      & $g\prime$,$r\prime$,$i\prime$,$z\prime_s$ & 8,8,7,8     \\
2024-01-13 01:22 & 2024-01-13 04:49 & c         & LCO1m/Sinistro@TO   & $r\prime$                                 & 20          \\
2024-01-24 06:37 & 2024-01-24 10:17 & d         & LCO2m/MuSCAT3@HO    & $g\prime$,$r\prime$,$i\prime$,$z\prime_s$ & 8,8,12,20   \\
2024-02-09 10:11 & 2024-02-09 13:04 & b         & Koyama1.3m/ADLER    & $z_s$                                     & 120         \\
2024-02-23 07:18 & 2024-02-23 08:58 & c         & LCO2m/MuSCAT3@HO    & $g\prime$,$r\prime$,$i\prime$,$z\prime_s$ & 8,8,12,20   \\
2024-03-04 20:00 & 2024-03-04 21:24 & b         & LCO1m/Sinistro@TO   & $r\prime$                                 & 20          \\
\botrule
\end{tabular*}
\footnotetext{CTIO = Cerro Tololo Inter-American Observatory in District IV, Chile \\
HO = Haleakala Observatory on Maui, Hawaii \\
MO = McDonald Observatory at Fort Davis, Texas \\
SSO = Siding Spring Observatory in New South Wales, Australia \\
TO = Teide Observatory on Tenerife in the Canary Islands, Spain}
\end{minipage}
\end{center}
\end{table}

\clearpage

\begin{table}[ht]
\tiny
\begin{center}
\begin{minipage}{\textwidth}
\caption{\sname measured transit times and 1$\sigma$ uncertainties. These data are also available as a separate machine-readable file from the publisher's website at \url{https://doi.org/10.1038/s41586-025-09840-z}. BKJD refers to BJD$-$2454833.}\label{sitab:2}
\begin{tabular*}{\textwidth}{*{4}{@{\hskip 5pt}l@{\hskip 5pt}}}
\toprule
Planet & Epoch & Transit midpoint (BKJD) & Facility \\
\midrule
c &      0 & $2231.2832 \pm 0.0020$ &                                   \ktwo \\
c &      1 & $2239.5327 \pm 0.0045$ &                                   \ktwo \\
c &      2 & $2247.7723 \pm 0.0098$ &                                   \ktwo \\
c &      3 & $2256.0350 \pm 0.0058$ &                                   \ktwo \\
c &      4 & $2264.2580 \pm 0.0064$ &                                   \ktwo \\
c &      5 & $2272.5311 \pm 0.0018$ &                                   \ktwo \\
c &      6 & $2280.7972 \pm 0.0067$ &                                   \ktwo \\
c &      7 & $2289.0226 \pm 0.0067$ &                                   \ktwo \\
c &      8 & $2297.2695 \pm 0.0034$ &                                   \ktwo \\
c &    210 & $3963.6070 \pm 0.0105$ &                                    TMMT \\
c &    211 & $3971.8525 \pm 0.0030$ &                      HDI/KeplerCam/TMMT \\
c &    214 & $3996.5942 \pm 0.0077$ &                                    TMMT \\
c &    216 & $4013.0976 \pm 0.0013$ &                                \spitzer \\
c &    218 & $4029.5995 \pm 0.0023$ &                                     APO \\
c &    222 & $4062.5848 \pm 0.0021$ &                                 APO/HDI \\
c &    293 & $4648.1657 \pm 0.0047$ &                                   \tess \\
c &    294 & $4656.4005 \pm 0.0074$ &                                   \tess \\
c &    296 & $4672.9002 \pm 0.0045$ &                                   \tess \\
c &    298 & $4689.3978 \pm 0.0069$ &                                   \tess \\
c &    339 & $5027.6711 \pm 0.0049$ &                          LCO1m$\times$1 \\
c &    351 & $5126.6737 \pm 0.0046$ &                          LCO1m$\times$3 \\
c &    383 & $5390.7061 \pm 0.0028$ &                  MuSCAT2/LCO1m$\times$4 \\
c &    384 & $5398.9600 \pm 0.0020$ &                          LCO1m$\times$2 \\
c &    385 & $5407.2092 \pm 0.0017$ &                                 MuSCAT3 \\
c &    387 & $5423.7178 \pm 0.0023$ &                                 MuSCAT2 \\
c &    388 & $5431.9562 \pm 0.0006$ &                                 MuSCAT3 \\
c &    390 & $5448.4547 \pm 0.0017$ &                                 MuSCAT2 \\
c &    391 & $5456.7008 \pm 0.0015$ &                       MuSCAT2/KeplerCam \\
c &    394 & $5481.4698 \pm 0.0027$ &                                 MuSCAT2 \\
c &    395 & $5489.7090 \pm 0.0049$ &                          LCO1m$\times$1 \\
c &    400 & $5530.9536 \pm 0.0021$ &                                 MuSCAT3 \\
d &      0 & $2239.3899 \pm 0.0039$ &                                   \ktwo \\
d &      1 & $2251.7913 \pm 0.0030$ &                                   \ktwo \\
d &      2 & $2264.2094 \pm 0.0069$ &                                   \ktwo \\
d &      3 & $2276.5991 \pm 0.0035$ &                                   \ktwo \\
d &      4 & $2289.0220 \pm 0.0061$ &                                   \ktwo \\
d &    143 & $4012.7974 \pm 0.0022$ &                                \spitzer \\
d &    166 & $4298.0983 \pm 0.0046$ &                          LCO1m$\times$1 \\
d &    171 & $4360.1189 \pm 0.0026$ &                                  Koyama \\
d &    194 & $4645.4116 \pm 0.0027$ &                                   \tess \\
d &    195 & $4657.8171 \pm 0.0033$ &                                   \tess \\
d &    196 & $4670.2131 \pm 0.0027$ &                                   \tess \\
d &    197 & $4682.6064 \pm 0.0040$ &                                   \tess \\
d &    227 & $5054.5934 \pm 0.0011$ &                                 MuSCAT2 \\
d &    228 & $5066.9944 \pm 0.0013$ &                                 MuSCAT3 \\
d &    229 & $5079.3889 \pm 0.0014$ &                  MuSCAT2/LCO1m$\times$1 \\
d &    232 & $5116.5786 \pm 0.0042$ &                          LCO1m$\times$2 \\
d &    235 & $5153.7473 \pm 0.0042$ &                          LCO1m$\times$2 \\
d &    255 & $5401.7488 \pm 0.0017$ &                          LCO1m$\times$3 \\
d &    258 & $5438.9578 \pm 0.0016$ &                                 MuSCAT3 \\
d &    260 & $5463.7558 \pm 0.0040$ &                               KeplerCam \\
d &    261 & $5476.1547 \pm 0.0028$ &                          LCO1m$\times$1 \\
d &    263 & $5500.9653 \pm 0.0022$ &                                 MuSCAT3 \\
b &      0 & $2234.0484 \pm 0.0020$ &                                   \ktwo \\
b &      1 & $2258.1898 \pm 0.0007$ &                                   \ktwo \\
b &      2 & $2282.3254 \pm 0.0030$ &                                   \ktwo \\
b &     65 & $3803.2404 \pm 0.0015$ &                                \spitzer \\
b &     74 & $4020.5009 \pm 0.0005$ &                            \spitzer/APO \\
b &     86 & $4310.1640 \pm 0.0007$ &                                  MuSCAT \\
b &    100 & $4648.0886 \pm 0.0018$ &                                   \tess \\
b &    101 & $4672.2288 \pm 0.0015$ &                                   \tess \\
b &    117 & $5058.4280 \pm 0.0055$ &                        LCO40cm$\times$1 \\
b &    118 & $5082.5685 \pm 0.0018$ &                          LCO1m$\times$3 \\
b &    119 & $5106.7201 \pm 0.0041$ &                          LCO1m$\times$1 \\
b &    122 & $5179.1312 \pm 0.0024$ &                                  Koyama \\
b &    128 & $5323.9664 \pm 0.0020$ &                                 MuSCAT3 \\
b &    129 & $5348.1105 \pm 0.0043$ &                          LCO1m$\times$2 \\
b &    134 & $5468.8072 \pm 0.0015$ &                          LCO1m$\times$2 \\
b &    136 & $5517.0876 \pm 0.0020$ &                                  Koyama \\
b &    137 & $5541.2227 \pm 0.0041$ &                          LCO1m$\times$2 \\
e &      0 & $2263.6229 \pm 0.0025$ &                                   \ktwo \\
e &     49 & $4648.7983 \pm 0.0013$ &                                   \tess \\
e &     57 & $5038.2516 \pm 0.0027$ & MuSCAT3/LCO1m$\times$1/LCO40cm$\times$1 \\
e &     59 & $5135.6100 \pm 0.0033$ &                          LCO1m$\times$2 \\
e &     65 & $5427.6810 \pm 0.0025$ &                  MuSCAT3/LCO1m$\times$2 \\
e &     66 & $5476.3465 \pm 0.0036$ &                          LCO1m$\times$1 \\
\botrule
\end{tabular*}
\footnotetext{BKJD refers to BJD$-$2454833.}
\end{minipage}
\end{center}
\end{table}

\clearpage

\begin{table}[ht]
\small
\begin{center}
\begin{minipage}{\textwidth}
\small
\caption{Posteriors from the multi-harmonic fit.}\label{sitab:3}
\begin{tabular}{LRr}
\toprule
\mathrm{Parameter} &   \mathrm{Value}  & Unit  \\
\midrule
\isr{P}{c}    & \val{ttv-per-c-mean}\pm\val{ttv-per-c-std}   & days \\ 
\isr{P}{d}    & \val{ttv-per-d-mean}\pm\val{ttv-per-d-std}   & days \\ 
\isr{P}{b}    & \val{ttv-per-b-mean}\pm\val{ttv-per-b-std}   & days \\ 
\isr{P}{e}    & \val{ttv-per-e-mean}\pm\val{ttv-per-e-std}   & days \\ 
\isr{t}{c,0}  & \val{ttv-t0-c-mean}\pm\val{ttv-t0-c-std}     & BKJD \\ 
\isr{t}{d,0}  & \val{ttv-t0-d-mean}\pm\val{ttv-t0-d-std}     & BKJD \\ 
\isr{t}{b,0}  & \val{ttv-t0-b-mean}\pm\val{ttv-t0-b-std}     & BKJD \\ 
\isr{t}{e,0}  & \val{ttv-t0-e-mean}\pm\val{ttv-t0-e-std}     & BKJD \\ 
\isr{A}{cd}   & \val{ttv-A-cd-mean}\pm\val{ttv-A-cd-std}     & days \\
\isr{A}{dc}   & \val{ttv-A-dc-mean}\pm\val{ttv-A-dc-std}     & days \\
\isr{A}{be}   & \val{ttv-A-be-mean}\pm\val{ttv-A-be-std}     & days \\
\isr{A}{eb}   & \val{ttv-A-eb-mean}\pm\val{ttv-A-eb-std}     & days \\
\isr{P}{cd}   & \val{ttv-per-cd-mean}\pm\val{ttv-per-cd-std} & days \\
\isr{P}{be}   & \val{ttv-per-be-mean}\pm\val{ttv-per-be-std} & days \\
\hline
\end{tabular}
\end{minipage}
\end{center}
\end{table}

\end{document}